\documentclass[aps,groupedaddress,11pt,superscriptaddress, longbibliography,floatfix,reprint]{revtex4-2}

\usepackage{amsmath}
\usepackage{amssymb}
\usepackage{graphicx}
\usepackage{hyperref}
\usepackage{bm}
\usepackage[dvipsnames]{xcolor}
\usepackage{setspace}
\usepackage{algpseudocode}
\makeatletter
\newcounter{algorithm}
\renewcommand\thealgorithm{\arabic{algorithm}}
\newenvironment{algorithm}[1][]{%
  \par\noindent\rule{\linewidth}{0.4pt}\par\nopagebreak[4]%
  \let\orig@caption\caption
  \renewcommand{\caption}[1]{%
    \par\refstepcounter{algorithm}%
    \medskip\noindent\textbf{Algorithm~\thealgorithm:} ##1\par\nopagebreak[4]\smallskip%
  }%
}{%
  \let\caption\orig@caption
  \par\noindent\rule{\linewidth}{0.4pt}\par\medskip%
}
\makeatother

\newcommand{\red}[1]{{\color{red} #1}}
\newcommand{\luke}[1]{{\color{blue}LE: #1}}
\newcommand{\henry}[1]{{\color{magenta}HM: #1}}

\DeclareMathOperator*{\argmax}{argmax} 

\hypersetup{colorlinks=true, linkcolor=blue!50!black, urlcolor=blue!50!black, citecolor=blue!50!black}

\newcommand{\dd}{\mathrm{d}}
\newcommand{\ee}{\mathrm{e}}

\newcommand{\E}{\mathbb{E}}
\newcommand{\Var}{\mathrm{Var}}
\newcommand{\Cov}{\mathrm{Cov}}

\newcommand{\KL}{D_{\mathrm{KL}}}
\newcommand{\Iay}{I(\alpha; Y)}
\newcommand{\Iaz}{I(\alpha; Z)}
\newcommand{\Iac}{I(\alpha; C)}
\newcommand{\Iacy}{I(\alpha; C \mid Y)}
\newcommand{\Iazy}{I(\alpha; Z \mid Y)}
\newcommand{\IayT}{I(\alpha; Y \mid \Theta)}

\newcommand{\Fyeta}{\bm{F}_{y|\eta}}
\newcommand{\Fyalpha}{\bm{F}_{y|\alpha}}
\newcommand{\FyalphaT}{\bm{F}_{y|\alpha,\theta}}

\DeclareMathOperator{\diag}{diag}
\DeclareMathOperator{\pdet}{pdet}

\newcommand{\Dir}{\mathrm{Dir}}
\newcommand{\Cat}{\mathrm{Cat}}
\newcommand{\Mult}{\mathrm{Mult}}

\newcommand{\V}{\bm{V}}
\newcommand{\J}{\bm{J}}
\newcommand{\R}{\bm{R}}
\newcommand{\U}{\bm{U}}
\newcommand{\D}{\bm{D}}
\newcommand{\I}{\bm{I}}
\newcommand{\B}{\bm{B}}

\newcommand{\ones}{\bm{1}}

\newcommand{\eps}{\varepsilon}

\newcommand{\order}{\mathcal{O}}

\newcommand{\SNR}{\mathrm{SNR}}
\newcommand{\signoise}{\sigma_{\mathrm{noise}}}
\newcommand{\sigsignal}{\sigma_{\mathrm{signal}}}

\newcommand{\avg}[1]{\left\langle #1\right\rangle}
\newcommand{\Fceta}{\bm{F}_{c|\eta}}
\newcommand{\Fcalpha}{\bm{F}_{c|\alpha}}
\newcommand{\Fyalphatilde}{\widetilde{\bm F}_{y|\alpha}}
\newcommand{\HI}{\bm{H}_I}
\newcommand{\deltamin}{\delta_{\mathrm{min}}}
\newcommand{\Mmax}{M_{\mathrm{max}}}
\newcommand{\DirMult}{\mathrm{Dir-Mult}}
\newcommand{\balpha}{\bm{\alpha}}
\newcommand{\bbeta}{\bm{\eta}}
\newcommand{\rr}{\bm{r}}
\newcommand{\s}{\bm{s}}
\newcommand{\sigatom}{\sigma_{\mathrm{bead}}}
\DeclareMathOperator{\Tr}{Tr}
\DeclareMathOperator{\sech}{sech}

\begin{document}
    \title{Measurement-limited learning of conformational heterogeneity in cryo-electron microscopy}

	\author{Henry H. Mattingly}
	\email{hmattingly@flatironinstitute.org}
	\affiliation{Center for Computational Biology, Flatiron Institute, New York, NY, USA}
    
	\author{Luke Evans}
	\affiliation{Center for Computational Mathematics, Flatiron Institute, New York, NY, USA}
    
	\author{Pilar Cossio}
    \affiliation{Center for Computational Biology, Flatiron Institute, New York, NY, USA}
	\affiliation{Center for Computational Mathematics, Flatiron Institute, New York, NY, USA}
	
	
	\begin{abstract} 
        Cryogenic electron microscopy images sample individual biomolecules from their conformational landscapes, offering a route to infer the distributions underlying molecular mechanisms. However, because images are indirect measurements, they limit which features of an underlying landscape are statistically identifiable. In ensemble reweighting, this problem appears as a choice of resolution: conformational space is discretized into representative structures whose population weights are inferred from images. Adding structures increases nominal resolution, but nearby conformations may generate overlapping image distributions and indistinguishable weights. Here, we develop an information-theoretic framework that selects representative conformations by maximizing mutual information between ensemble weights and images under a probabilistic forward model. Analytically, we show in a one-dimensional Gaussian model that measurement noise sets the optimal spacing. Applied to molecular conformations sampled from simulation, the framework constructs near-optimal ensembles that span heterogeneity while avoiding redundancy. Thus, the measurement process induces a maximally learnable coarse graining of conformation space.
	\end{abstract}
	
	\maketitle

	\section{Introduction}

    Biomolecules dynamically explore complex energy landscapes to perform biological functions \cite{frauenfelder_energy_1991, boehr_role_2009}. Many of these functions emerge from transitions among metastable conformations, such as in ion-channel gating, allostery, and ligand- or environment-induced RNA rearrangements \cite{smock_sending_2009, astore_protein_2024, al-hashimi_rna_2008, herschlag_story_2018, bonilla_promise_2022, bonilla_challenges_2024}. Defining these conformational states and their thermodynamic probabilities is therefore central to understanding molecular mechanisms. 
    
    Experimentally, conformational distributions must be inferred from finite, noisy measurements. This raises the question of which features of the underlying landscape are learnable from a given experiment. Since any physical measurement process obscures differences among conformational ensembles, there is a statistical limit on our ability to resolve heterogeneity. Therefore, while map resolution in structural biology is the spatial scale of resolvable structural features in a 3D volume, finite measurements induce an analogous concept of \textit{conformational resolution}, the scale at which the conformational landscapes can reliably be inferred.
    
    Single-particle cryo-electron microscopy (cryo-EM) provides a high-resolution window into conformation landscapes and a natural setting for studying the limits of conformational resolution. Images are collected of individual biomolecules in varying conformations, which are nearly drawn from their equilibrium distribution owing to rapid vitrification \cite{clark_cooling_2026}. These images can reach near-atomic spatial resolution after averaging many particles \cite{kuhlbrandt_resolution_2014, cheng_single-particle_2015}, but individual particles have low signal-to-noise ratio \cite{scheres_relion_2012, cossio_bayesian_2013, singer_computational_2020}. As a result, conventional reconstruction often relies on averaging particles into a small number of consensus maps \cite{bendory_single-particle_2020, singer_computational_2020}, potentially discarding thermodynamic information about heterogeneous molecular ensembles that is present in the single-particle images.

\begin{figure*}
		\centering
		\includegraphics[width=\textwidth]{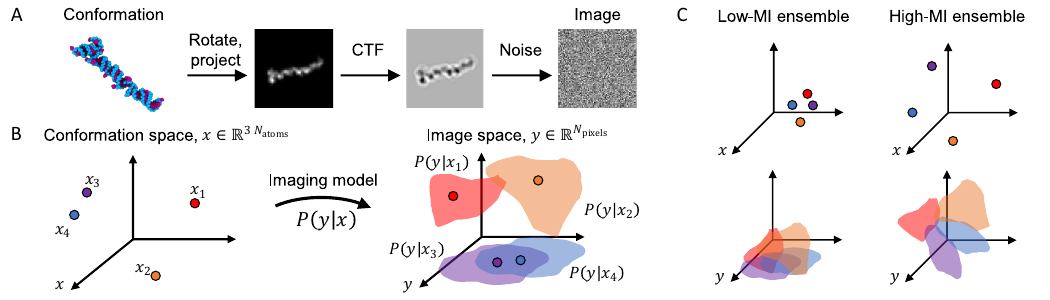}
		\caption{A) Schematic of the cryo-EM imaging process that probabilistically maps a conformation $x$ to an image $y$ by modeling the unknown rotation, projection, center translation, contrast transfer function (CTF) and noise. B) Each molecular conformation $x$ generates a distribution of images $P(y \mid x)$ under the forward model. Distinguishability is defined in image space $y$: nearby conformations often produce overlapping image distributions, and even distant conformations can produce indistinguishable images.
		C) Maximizing mutual information favors ensembles of conformations whose image distributions maximally cover image space with minimal redundancy.}
		\label{fig:schematic}
	\end{figure*}

    Ensemble reweighting offers a route to recovering this information by combining candidate molecular structures with experimental data to infer conformational populations \cite{bonomi_principles_2017, kofinger_efficient_2019, orioli_how_2020}. In this approach, conformational space is discretized into a finite set of representative structures whose statistical weights are inferred from data. Recent work has extended this idea to cryo-EM, where the data consist of noisy single-particle images \cite{tang_ensemble_2023, tang_conformational_2023}. A probabilistic forward model maps each representative to a distribution of possible images by accounting for latent imaging variables and experimental noise, including unknown orientation and center translation, microscope effects, pixelization, and noise (Figure \ref{fig:schematic}A) \cite{scheres_relion_2012, cossio_bayesian_2013, singer_computational_2020, tang_ensemble_2023}. The observed image distribution is therefore modeled as a mixture whose components are the image distributions generated by the representative structures, with the component weights as unknown parameters.

    This formulation leaves open a fundamental representation problem: how should conformational space be discretized? In practice, candidate structures are typically drawn from molecular dynamics, Monte Carlo sampling, or generative modeling \cite{noe_machine_2020, zhong_cryodrgn_2021}, but which and how many structures to include in the ensemble has generally been chosen heuristically. These methods can generate many diverse conformations, and including more of them in the discretization increases the nominal resolution with which the conformational landscape can be represented. However, if nearby conformations generate highly overlapping image distributions, then redistributing weight among them produces little change in the predicted observations.

    This limitation is a form of parameter unidentifiability in mixture models. Classical results show that mixture models become singular when their components produce identical or nearly identical observations, causing parameters such as mixture weights to become difficult or impossible to estimate \cite{vardi_image_1993, lindsay_mixture_1995, watanabe_almost_2007, watanabe_algebraic_2009}. In the cryo-EM setting, structurally similar representatives can therefore introduce nearly flat directions in the likelihood function, along which parameters are unconstrained. This motivates choosing the conformational representation to capture as much heterogeneity as possible while maintaining identifiability of the associated weights.
    
    A natural objective for this tradeoff is the mutual information (MI) between ensemble weights and cryo-EM data expected under the forward model \cite{shannon_mathematical_1948}. In this context, MI quantifies the expected reduction in uncertainty about unknown parameters after observing data \cite{lindley_measure_1956}. Previous work has shown that maximizing MI favors representations supported on distinguishable parameter values \cite{bernardo_reference_1979, berger_formal_2009, mattingly_maximizing_2018, abbott_far_2023}, making it well suited to balancing conformational resolution against parameter identifiability. Because this objective is defined through the forward model and prior assumptions, it can be evaluated before any experimental images are observed.
    
    Here, we develop an information-theoretic framework for measurement-limited learning of conformational heterogeneity in cryo-EM. We maximize the mutual information between ensemble weights and images to select finite conformational representations whose populations are expected to be most learnable from data. Analytically, we show that overlap between image distributions shrinks Fisher information eigenvalues, and that in a one-dimensional Gaussian model the optimal spacing of representative structures is set by the measurement noise. We then compute the MI in a high-dimensional cryo-EM setting and construct near-optimal ensembles from molecular simulations. Notably, since this analysis relies solely on the forward model, its parameters, and their prior distributions, the conformational representation can be chosen before observing experimental data. Our results establish a measurement-induced coarse-graining of conformational heterogeneity, linking the physical process of image formation to the learnable degrees of freedom of an equilibrium ensemble.

	\section{Problem setup}

    We model a biomolecule's equilibrium conformational distribution by discretizing conformational space into representative structures, $x \in \mathbb{R}^{3N_{atom}}$ \cite{tang_ensemble_2023}. A discretization consists of a set of $M$ representative structures $X=\{x_m\}_{m=1}^M$ with weights $\alpha = (\alpha_1,\dots,\alpha_M)$ satisfying $\alpha_m \ge 0$ and $\sum_{m=1}^M \alpha_m = 1$. These together define a model probability distribution over conformations (Appendix~\ref{SIsec:problem}),
    \begin{equation}
        P(x \mid \alpha) =\sum_{m=1}^M \delta(x-x_m) \, \alpha_m~.
        \label{eq:ensemble}
    \end{equation}
    Since we observe images but not the conformations that generated them, the likelihood of a single image $y$, $P(y \mid \alpha) = \int P(y \mid x)  \, P(x \mid \alpha) \, \dd x$, has the following mixture form
    \begin{equation}
        P(y \mid \alpha) = \sum_{m=1}^M P(y \mid x_m) \, \alpha_m~,
        \label{eq:mixture_likelihood}
    \end{equation}
    where we have used Eq.~\eqref{eq:ensemble}, and $P(y \mid x)$ encodes the full probabilistic process mapping structures to images (Figure \ref{fig:schematic}A). A dataset $Y = \{y_i\}_{i=1}^N$ consists of $N$ i.i.d samples from $P(y \mid \alpha)$ for some unknown parameters $\alpha$. 

    We take a Bayesian approach, where data are used to compute the posterior distribution over parameters, 
    \begin{equation}
        P(\alpha \mid Y) = \frac{P(Y \mid \alpha) \, P(\alpha)}{P(Y)},
    \end{equation}
    where $P(\alpha)$ is the prior distribution over the weights and $P(Y) = \int P(Y \mid \alpha) \, P(\alpha) \, \dd \alpha$ is the model distribution of image data implied by the prior and the forward model. We use a symmetric Dirichlet prior with concentration parameter $\beta$.

    The MI between parameters and data, $\Iay$, quantifies the expected information gain about the parameters upon observing data
    \begin{equation}
    	\Iay
    	=
    	H(\alpha)
    	-
    	\E_{Y}[H(\alpha\mid Y)]~,
    	\label{eq:MI_definition}
    \end{equation}
    where 
    \begin{equation}
        H(\alpha) = - \int P(\alpha) \, \log P(\alpha) \, \dd \alpha 
    \end{equation}
    is the entropy of the prior distribution and 
    \begin{equation}
        H(\alpha\mid Y) = - \int P(\alpha \mid Y) \, \log P(\alpha \mid Y) \, \dd \alpha
    \end{equation}
    is the entropy of the posterior distribution for some fixed data $Y$. Then, the expectation $\E_{Y}[H(\alpha\mid Y)]$ in Eq.~\eqref{eq:MI_definition} is taken over image data $Y$ drawn from the model distribution $P(Y)$ (i.e., simulated images). Thus, $\Iay$ quantifies how much our uncertainty about $\alpha$ is reduced, on average, by observing data $Y$. Importantly, $\Iay$ can be evaluated before observing data, allowing us to quantify the learnability of the model \textit{a priori}.

    Our objective is to determine a discrete set of representative structures—both the number of components $M$ and the identities of the structures $\{x_m\}$— that maximizes $\Iay$. Increasing $M$ improves the resolution of the model, but it also introduces additional parameters that are only useful if the corresponding structures remain distinguishable under the forward model. Thus, the problem is to construct an ensemble that balances model resolution against identifiability (Figure \ref{fig:schematic}C).

	\section{Image overlap makes ensemble weights unidentifiable}

    Directly observing molecular conformations would simplify inference of their equilibrium weights. Cryo-EM images make this problem harder because they provide only noisy, indirect evidence about which conformations generated the data. Under the discretized model, each image is generated by sampling a latent component $z\in\{1,\dots,M\}$ with probability $\alpha_z$, and then producing an image from $P(y\mid x_z)$. Observing the latent components $Z=\{z_i\}_{i=1}^N$ would provide information $\Iaz$ about the weights; the observed images $Y$ contain only the part of this information that survives the imaging process. Since $\alpha\to Z\to Y$, the chain rule of mutual information gives
    \begin{equation}
        \Iay = \Iaz-\Iazy ~.
    \end{equation}
    The second term is the information about $\alpha$ that remains in the latent assignments after observing the images, and therefore quantifies assignment ambiguity introduced by the imaging process (Appendix~\ref{SIsec:MI_def}). If image distributions do not overlap, each image identifies its component, $\Iazy=0$, and weights can be learned by particle counting \cite{evans_counting_2025}. Overlap among image distributions makes assignments ambiguous and reduces $\Iay$ (Fig.~\ref{fig:schematic}B).

    To quantify this loss, we use a Gaussian approximation to $\Iay$ (Appendix~\ref{SIsec:largeN_MI}). With Dirichlet prior covariance $\Sigma_\alpha$, the large-$N$ posterior precision is
	\begin{equation}
		\Sigma_{\alpha|Y}^{-1} \approx N \, \Fyalpha(\alpha) + \Sigma_\alpha^{-1}~,
	\end{equation}
	where $\Fyalpha(\alpha) \in \mathbb{R}^{M \times M}$ is the Fisher information matrix \cite{amari_methods_2000},
	\begin{equation}
		\Fyalpha(\alpha) = \E_{y|\alpha}\!\left[\nabla_{\alpha} \log P(y\mid\alpha) \, (\nabla_{\alpha} \log P(y\mid\alpha))^\top \right]~.
		\label{eq:Fisher_def}
	\end{equation}
    The eigenvalues of $\Fyalpha(\alpha)$, restricted to directions that preserve $\sum_m\alpha_m=1$ (Appendix~\ref{SIsubsec:Gaussian_approx}), quantify how precisely nearby changes in the weights can be inferred from images. Denoting these eigenvalues as $\lambda_k(\alpha)$, $k=1,\dots,M-1$, then under the Gaussian approximation, the mutual information is
	\begin{equation}
		\Iay \approx \frac{1}{2} \sum_{k=1}^{M-1} \E_\alpha\!\left[\log\!\left(1 + \frac{N \lambda_k(\alpha)}{M(M\beta + 1)} \right)\right]~.
		\label{eq:Iay_Gaussian}
	\end{equation}

    \begin{figure*}
		\centering
		\includegraphics[width=\textwidth]{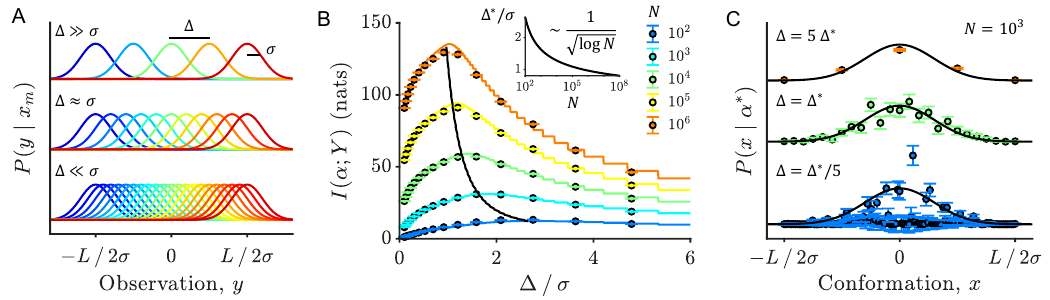}
		\caption{
        A) Schematic of 1D Gaussian toy system. Representative structures $x_m\in [-L/2\sigma,L/2\sigma]$ and data $Y$ are both scalars, with $y \sim \mathcal N(x,\sigma^2)$. Shown are $P(y \mid x_m)$ for varying $m$ (colors) with different spacing $\Delta$ (decreasing from top to bottom).
		B) Mutual information $\Iay$ versus spacing $\Delta / \sigma$ for varying $N$ (colors). Markers: numerical results for $\Iay$ using the Gaussian approximation in Eq.~\eqref{eq:Iay_Gaussian}. Error bars standard errors of the mean over 5 replicates. Solid lines: theoretical expression in Eq.~\eqref{eq:MI_1D}. Black: Analytical prediction for the optimal spacing $\Delta^*/\sigma$ (Eq.~\eqref{eq:Delta_opt}) and the optimal value of $I(\alpha;Y)$. In (B) and (C), $L/\sigma = 40$.
        Inset: Predicted scaling of $\Delta^*/\sigma$ with $N$ from Eq.~\eqref{eq:Delta_opt}.
        C) Parameter inferences for varying ensembles. Black lines: true distribution from which $N=10^3$ data points $Y$ were sampled. Markers: Posterior maximum (MAP)  distributions $P(x\mid \alpha^*)$ versus $x$, using ensembles with varying spacing $\Delta$ and the same data $Y$. Error bars are posterior standard deviations of the $\alpha_m$. The optimal spacing provides the best balance between ensemble resolution and parameter identifiability. 
        }
		\label{fig:1D_model}
	\end{figure*}

    The connection between image overlap and Fisher information is made explicit by the responsibility vectors, $r\in \mathbb{R}^M$ (Appendix~\ref{SIsubsec:Fisher_responsibilities}). Element $r_m$ represents the probability that structure $m$ generated image $y$, given $\alpha$,
	\begin{equation}
		r_m(y;\alpha) \equiv P(z=m \mid y,\alpha) = \frac{P(y \mid x_m) \, \alpha_m}{\sum_j P(y \mid x_j) \, \alpha_j} ~.
		\label{eq:responsibilities}
	\end{equation}
	In terms of these responsibilities, the Fisher information matrix has entries
	\begin{equation}
        \left( \Fyalpha(\alpha) \right)_{m,m'} = \E_{y|\alpha}\!\left[\frac{r_m(y;\alpha)}{\alpha_m} \, \frac{r_{m'}(y;\alpha)}{\alpha_{m'}}\right]~,
		\label{eq:Fisher_entries}
	\end{equation}
    before restricting to directions within the simplex. The expectation in Eq.~\eqref{eq:Fisher_entries} quantifies the overlap between $P(y \mid x_m)$ and $P(y \mid x_{m'})$ and thus the ambiguity in assigning data to components. Overlapping components tend to have similar responsibility entries, reducing Fisher eigenvalues along directions that redistribute weight among those components (Section III.C), thus reducing the information gained from data $\Iay$ (Eq.~\eqref{eq:Iay_Gaussian}).

    Thus, increasing the number of structures $M$ has two competing effects. It adds parameter directions that could be learned and terms to the sum in Eq.~\eqref{eq:Iay_Gaussian}. At the same time, overlap among their image distributions reduces the identifiability of those directions. Finite data must also be distributed among more parameters, reducing the statistical power available for each one. Together, these effects indicate that additional structures only increase $\Iay$ when the new parameter directions are identifiable---namely, when $N \lambda_k(\alpha) \gtrsim M(M\beta+1)$. The optimal discretization balances this trade off, favoring ensembles whose components span the space of observable images while minimizing redundancy under the forward model.

	\section{Measurement noise sets an optimal spacing in 1D}

	To show how a noisy measurement induces a learnable coarse-graining, we first consider a minimal measurement model: a continuous one-dimensional conformational coordinate corrupted with Gaussian noise (Fig.~\ref{fig:1D_model}A, Appendix~\ref{SIsec:1D_model}). We consider conformations $x \in [-L/2,L/2]$ discretized into evenly spaced components with spacing $\Delta$, giving $M \approx 1 + L/\Delta$ (Figure \ref{fig:1D_model}A). The forward model is
	\begin{equation}
		P(y \mid x_m) = \mathcal{N}(y; x_m, \sigma^2) ~,
		\label{eq:Qm_1D}
	\end{equation}
	with $x_m = (m-1) \, \Delta - L/2$ and $m = 1, \dots, M$. 

	In this minimal model, the Fisher information can be diagonalized analytically when $M$ is large ($L \gg \sigma$). We Taylor expand the argument of the expectation over $\alpha$ in Eq.~\eqref{eq:Iay_Gaussian} about the prior mean, $\bar \alpha = \ones / M$. The leading-order approximation to $\Iay$ thus replaces the expectation over $\alpha$ by evaluating its argument at $\alpha = \bar\alpha$. In Appendix~\ref{SIsubsec:1D_Taylor_correction}, we show that the correction to this approximation is smaller by a factor of order $\order(M^{-2})$, which is small for large $M$. In the same regime, boundaries have small effects and the Fisher information matrix is nearly circulant, allowing its eigenvalues to be derived analytically. For moderate $\delta \equiv \Delta/\sigma$, the $M-1$ eigenvalues are
	\begin{equation}
		\lambda_k \approx M \left(e^{-(2\pi k/M)^2/\delta^2} + e^{-\left(2\pi (1 - k/M)\right)^2/\delta^2}\right)~.
		\label{eq:Fisher_eigs_1D}
	\end{equation}
	The smallest eigenvalue $k=M/2$ corresponds to the “Nyquist” mode, $\phi_{M/2}(m)=\cos(\pi \, m)$, which alternates weight between neighboring components. High-frequency modes are least identifiable when component distributions overlap because they redistribute weight on length scales smaller than the measurement noise, producing little change in the observed data distribution.
	
	Using the symmetry $\lambda_k=\lambda_{M-k}$ and the observation that only one of the two exponentials in Eq.~\eqref{eq:Fisher_eigs_1D} dominates at each $k$, the mutual information is approximately
	\begin{equation}
		\begin{aligned}
			\Iay &\approx \sum_{k=1}^{M/2-1} \log\!\left(1 + \frac{N}{M \, \beta + 1} \, e^{-(2\pi k/M)^2/\delta^2} \right)
			\\
			&\qquad + \frac{1}{2} \log\!\left(1 + \frac{2\, N}{M \, \beta + 1} \, e^{-\pi^2/\delta^2}\right).
		\end{aligned}
		\label{eq:MI_1D}
	\end{equation}
	
	Figure \ref{fig:1D_model}B shows $\Iay$ computed numerically using the Gaussian approximation in Eq.~\eqref{eq:Iay_Gaussian} and the analytical expression in Eq.~\eqref{eq:MI_1D}, for varying $\delta$ and $N$. The analytical expression shows excellent agreement with the numerical results, especially for large $N$.

	To optimize $\Iay$, we balance the information gain from adding parameters against the information loss from reducing their identifiability. This yields
	\begin{equation}
		\Delta^* \approx \sigma \, \pi \, \left(\log\left(\frac{\sqrt{2} \, \pi \, N}{L/\sigma \, \beta \, e}\right)\right)^{-1/2}~.
		\label{eq:Delta_opt}
	\end{equation}
	Thus, $\Delta^*$ is largely set by the noise scale, $\sigma$: components should be separated so their data distributions are \textit{just distinguishable}. Although $\Delta^*$ decreases with $N$, this dependence is weak, $\Delta^* \sim (\log N)^{-1/2}$ (Figure \ref{fig:1D_model}B inset), indicating that discretization is governed primarily by measurement noise rather than sample size.

    We next inferred $\alpha$ for ensembles with different spacings $\Delta$, using Markov Chain Monte Carlo to sample each posterior for the same dataset of $N=10^3$ noisy observations from a true Gaussian distribution over $x$ (Appendix~\ref{SIsec:MCMC}). Figure~\ref{fig:1D_model}C shows the inferred distributions at the posterior maximum, $\alpha^*$. When $\Delta \gg \Delta^*$, the weights are well constrained but the grid is too coarse to resolve the distribution. When $\Delta \ll \Delta^*$, the MAP estimate becomes sparse, selecting only a subset of components, consistent with the self-regularizing behavior of nonparametric maximum likelihood estimators \cite{polyanskiy_self-regularizing_2020}. However, many sparse subsets have similar likelihoods, so the selected fine-scale components are not unique. The optimal spacing $\Delta^*$ gives the finest discretization whose weights are learnable from the data.

    \begin{figure*}
		\centering
		\includegraphics[width=\textwidth]{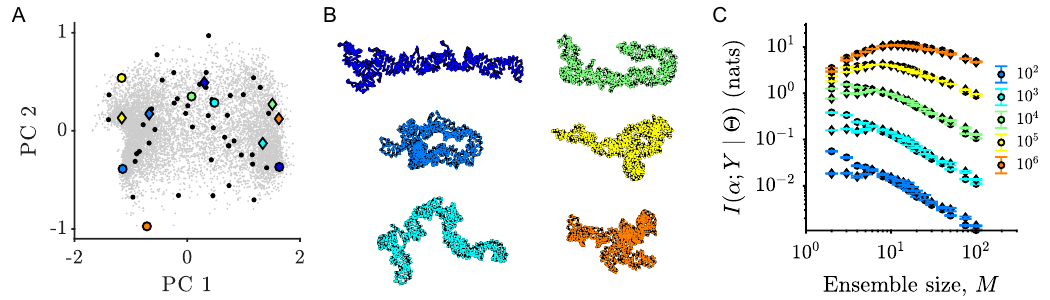}
		\caption{
        A) Conformational fluctuations of the RNA P4-P6 domain of the \textit{Tetrahymena thermophila} group I intron ribozyme was simulated in Martini. Shown are the projections of all frames onto the first two principal components, scaled to have unit variance. Gray: conformations from all simulation frames. Circles: the first six conformations selected by a greedy algorithm, from dark blue to orange. Black: the next 44 conformations selected by the farthest-point algorithm. Diamonds: the first six conformations selected by a hierarchical k-medoids algorithm.
        B) RNA conformations from the farthest-point ensemble corresponding to the colored circles in (A). The greedy algorithm first selects the most open conformation (dark blue), then the most closed one (medium blue).
		C) Numerical calculation of Gaussian mutual information $\Iay$ (Eq.~\eqref{eq:IayT_Gaussian}) versus ensemble size $M$ for varying sample sizes $N$ (colors). Circles: Farthest-point ensembles. Diamonds: k-medoids ensembles. Error bars are standard errors of 5 replicates. $\Iay$ only has a peak at finite $M$ once $N$ is large enough.
		}
		\label{fig:RNA_ensemble}
	\end{figure*}
	
	\section{Coarse-graining conformational samples into learnable ensembles}

    In the Gaussian model, the optimal spacing was set by measurement noise. For molecular cryo-EM, the corresponding notion of spacing is not a structural distance alone, but distinguishability under the full image-formation process. Unknown orientation, center translation, projection, microscope effects, pixelization, and noise all shape the image distribution generated by a conformation. Thus, conformations that are distinct in molecular coordinates can become statistically indistinguishable after passing through the forward model. We therefore asked whether mutual information can be used to select representative structures from a molecular simulation or generative model.

	The cryo-EM forward model consists of image-specific, known and unknown parameters. For example, an in-focus image in cryo-EM has no contrast \cite{frank_three-dimensional_2006}, so molecules are intentionally imaged out of focus with known defocus. Let $\theta_i$ denote known parameters for image $y_i$ and let $\phi_i$ denote unknown parameters, including rotation $R$, center translation $\bm{\tau}$, and noise variance $\sigma_{\mathrm{noise}}^2$. For a structure $x$, the likelihood conditioned on known parameters is obtained by marginalizing over the unknown ones:
	\begin{equation}
    	P(y\mid x,\theta)
    	=
    	\int P(y\mid x,\phi,\theta) \, P(\phi) \, \dd\phi ~.
    	\label{eq:Pyx_molecule}
	\end{equation}
	In our implementation, each structure is randomly rotated and translated; Gaussian densities are placed at atomic positions; these densities are projected onto a pixel grid and passed through a contrast transfer function; and each pixel is corrupted by Gaussian noise (Fig.~\ref{fig:schematic}A). These operations were performed using the cryo-EM image simulator in the cryoJAX Python package \cite{obrien_cryojax_2026}. We take $P(R)$ to be uniform, $P(\bm{\tau})$ uniform within a radius of 20 pixels from the image center, and $P(\sigma_{\mathrm{noise}}^2)$ log-uniform over a range corresponding to $\mathrm{SNR}\in[0.05,0.1]$ \cite{baxter_determination_2009,seitz_simulation_2019} (refer to Appendix~\ref{SIsec:noise_level} for how SNR was computed). 
    
    With this, the likelihood of image $y$ with weights $\alpha$, conditioned on $\theta$, is
	\begin{equation}
    	P(y\mid \alpha,\theta)
    	=
    	\sum_{m=1}^M P(y\mid x_m,\theta) \, \alpha_m ~.
	\end{equation}
    Since $\theta_i$ is known for each image, the relevant information is the conditional mutual information between $\alpha$ and images given the known parameters,
	\begin{equation}
	    \IayT
	    =
	    H(\alpha)-\E_{Y,\Theta}[H(\alpha\mid Y,\Theta)] ~,
	    \label{eq:MI_Theta_def}
	\end{equation}
	where $\Theta = \{\theta_1, \dots, \theta_n\}$. This in turn depends on the per-image Fisher information associated with the likelihood $P(y\mid \alpha,\theta)$, denoted $\FyalphaT(\alpha,\theta)$.
	For data sets with many images $N$, draws of $\theta_i$ across images self-average, so the relevant Fisher information is (Appendix~\ref{SIsec:conditional_MI})
	\begin{equation}
	    \bar{\bm F}_{y|\alpha}(\alpha)
	    =
	    \E_\theta\!\left[\FyalphaT (\alpha,\theta)\right]~,
	\end{equation}
	with eigenvalues $\bar\lambda_k(\alpha)$. Then, the Gaussian approximation of $\IayT$ is
	\begin{equation}
    	\IayT
    	\approx
    	\frac{1}{2}\sum_{k=1}^{M-1}
    	\E_\alpha\!\left[
    	\log\!\left(
    	1+\frac{N\bar\lambda_k(\alpha)}{M(M\beta+1)}
    	\right)
    	\right]~.
    	\label{eq:IayT_Gaussian}
	\end{equation}

    Eq.~\eqref{eq:IayT_Gaussian} assigns each potential conformational ensemble a scalar learnability score under the cryo-EM forward model. To apply this criterion to molecular conformations, we generated samples of the RNA P4-P6 domain of the \textit{Tetrahymena thermophila} group I intron ribozyme using coarse-grained MD simulations with the Martini2 force field in GROMACS \cite{kruger_self-splicing_1982, woodson_folding_2002, cech_ribozymes_2002, bonilla_cryo-em_2022, uusitalo_martini_2017, abraham_gromacs_2015} (Appendix~\ref{SIsec:mol_sim_data}). From these samples, we constructed nested ensembles using a greedy farthest-point algorithm \cite{eldar_farthest_1997}: starting from the structure farthest from the trajectory average, we iteratively add the conformation whose minimum RMSD to the current ensemble members is largest (Algorithm~\ref{SIalg:farthest-point_ensemble}, Appendix~\ref{SIsec:farthest-point_ensemble}). This yields ensembles that progressively cover conformational space while avoiding redundancy. As a comparison, we also constructed nested ensembles using hierarchical k-medoids clustering, which chooses representative conformations according to the density of the MD trajectory. Determining the optimal size $M^*$ then reduces to maximizing $\IayT$ over $M$.

    A PCA embedding of all simulation frames is shown in Fig.~\ref{fig:RNA_ensemble}A, with the first six selected structures from each ensemble marked by colored dots. Fig.~\ref{fig:RNA_ensemble}B shows these conformations for the farthest-point algorithm: the first two correspond to maximally open and closed states, while later selections capture intermediate states.

    Using the Gaussian approximation in Eq.~\eqref{eq:IayT_Gaussian}, we computed $\IayT$ for both ensembles as a function of ensemble size $M$, over a range considered previously \cite{tang_conformational_2023}, and number of images $N$ (Fig.~\ref{fig:RNA_ensemble}C; Appendix~\ref{SIsec:methods}). For small $N$, $\IayT$ saturates after only a few components, indicating that limited data support only the most distinguishable conformational variation---open and closed RNA conformations (Fig.~\ref{fig:RNA_ensemble}B). In the farthest-point ensembles, this variation is captured already at an ensemble with $M=2$ conformations, whereas the k-medoids ensembles flatten or peak at small finite $M$. Once the number of images is large, $N\gtrsim 10^4$, $\IayT$ is maximized at larger $M$, and the optimal ensemble size $M^*$ increases with $N$. For reference, a cryo-EM experiment typically collects $10^6$ particles, which may be reduced to $10^4$ images in the process of quality control. Thus, molecular simulations provide candidate heterogeneity, while the measurement model determines how much and which part of that heterogeneity can be learned from data.

	\section{Discussion}

    We have shown that the image formation process in cryo-EM defines a resolution limit on learnable conformational heterogeneity. Adding more structures refines the representation of conformational space, but it also introduces more ensemble weights to infer from the same data. When nearby structures generate overlapping image distributions, their weights compete to explain the same observations, creating poorly identifiable directions in parameter space. Mutual information captures this tradeoff and selects finite ensembles whose populations are most learnable under the forward model. Because distinguishability is set by the measurement process, the learnable conformational resolution depends on noise level, microscope parameters, nuisance-variable priors, sample size, and the molecule itself.

    This work connects ensemble reweighting to a broader principle of finite-data model selection: when data constrain only a subset of parameter directions, maximizing information favors lower-dimensional effective representations \cite{mattingly_maximizing_2018, lamont_correspondence_2019}. Here, the parameter directions are statistical weights over molecular conformations, and their distinguishability is set by the cryo-EM forward model. The same logic applies to other representations of conformational heterogeneity in cryo-EM, including neural-network maps from latent coordinates to molecular structures or latent atomic representations learned directly from images: additional degrees of freedom are learnable only when they introduce distinguishable changes in the image distribution.
    
    Our information-theoretic approach can also guide ensemble construction for ensemble-averaged measurements such as SAXS, NMR, or FRET \cite{bonomi_principles_2017, kofinger_efficient_2019, bottaro_integrating_2020, orioli_how_2020}. In those settings, parameter unidentifiability arises when distinct reweightings produce nearly identical averaged observables. Extending the framework would therefore mean selecting structures whose weights have distinguishable effects on the measured observables, an important direction for future work.

    Our criterion is complementary to maximum-entropy approaches commonly used in ensemble reweighting \cite{orioli_how_2020, borthakur_determining_2025}. MaxEnt addresses inference for a specified ensemble: how should weights be updated to match data while remaining close to a prior? In contrast, our framework addresses the model-design problem: which representative structures should be included before inference is performed? Thus, both approaches can be used in the same analysis pipeline. In fact, MaxEnt with soft constraints can be posed as a Bayesian inference problem with a specific choice of prior (Appendix~\ref{sec:maxent_bayes}).
    
    In this work, we used an RMSD-based farthest-point procedure to generate nested candidate ensembles before scoring them with the image-space MI objective. This emphasizes coverage of conformational variation while discarding density information from the molecular simulation. If the simulation or generative model provides a useful prior over conformational space, density-aware methods such as k-medoids may be preferable. Separately, future work could optimize representative structures more directly under the forward model \cite{silva-sanchez_cryo-electron_2026}, for example by combining MD with a repulsion force among ensemble members based on the overlap among $P(y \mid x)$ to find higher-MI ensembles.

    More broadly, this framework connects molecular-state definitions to the physics of measurement. Metastability, function, and molecular mechanisms define meaningful conformational states, but the experimental readout determines which features can be reliably assigned probabilities. By making this dependence explicit, our work provides a way to construct coarse-grained molecular ensembles at the conformational resolution induced by the measurement process.

    \paragraph{Code availability.} Code used to produce the results will be made available on GitHub. 

    \paragraph{Data availability.} The concatenated RNA molecular dynamics trajectories will be made available.

    \paragraph{Acknowledgements.}
    We thank F. Emil Thomasen and Matthew Leighton for comments on the manuscript. We also thank Steve Bonilla and the COMPASS and Biophysical Modeling groups at the Flatiron Institute for feedback at early stages of the project. The Flatiron Institute is a division of the Simons Foundation.
    
    \paragraph{Author contributions.} HHM conceptualized the work. HHM, LE, and PC designed the research. HHM performed the derivations, wrote the code, and ran the computations. PC provided the RNA simulation data. HHM, LE, and PC wrote and edited the manuscript.

	\appendix
	\newcommand{\combinedwithmain}{}
	\onecolumngrid
	\ifdefined\combinedwithmain\else
\documentclass[aps,groupedaddress,11pt,superscriptaddress, longbibliography,onecolumn]{revtex4-2}

\usepackage{amsmath,amssymb,bm}
\usepackage{graphicx}
\usepackage[dvipsnames]{xcolor}
\usepackage{setspace}

\usepackage{algpseudocode}
\makeatletter
\newcounter{algorithm}
\renewcommand\thealgorithm{\arabic{algorithm}}
\newenvironment{algorithm}[1][]{%
  \par\noindent\rule{\linewidth}{0.4pt}\par\nopagebreak[4]%
  \let\orig@caption\caption
  \renewcommand{\caption}[1]{%
    \par\refstepcounter{algorithm}%
    \medskip\noindent\textbf{Algorithm~\thealgorithm:} ##1\par\nopagebreak[4]\smallskip%
  }%
}{%
  \let\caption\orig@caption
  \par\noindent\rule{\linewidth}{0.4pt}\par\medskip%
}
\makeatother
\usepackage{hyperref}
\hypersetup{colorlinks=true, linkcolor=blue!50!black, urlcolor=blue!50!black, citecolor=blue!50!black}

\newcommand{\dd}{\mathrm d}
\newcommand{\E}{\mathbb E}
\newcommand{\Var}{\mathrm{Var}}
\newcommand{\Cov}{\mathrm{Cov}}
\newcommand{\avg}[1]{\left\langle #1\right\rangle}
\newcommand{\ones}{\bm{1}}
\newcommand{\Iay}{I(\alpha; Y)}
\newcommand{\IayT}{I(\alpha; Y \mid \Theta)}
\newcommand{\Iaz}{I(\alpha; Z)}
\newcommand{\Iac}{I(\alpha; C)}
\newcommand{\Iacy}{I(\alpha; C \mid Y)}
\newcommand{\Dir}{\mathrm{Dir}}
\newcommand{\Mult}{\mathrm{Mult}}
\newcommand{\DirMult}{\mathrm{Dir-Mult}}
\newcommand{\Cat}{\mathrm{Cat}}
\newcommand{\KL}{D_{\mathrm{KL}}}
\newcommand{\eps}{\varepsilon}
\newcommand{\D}{\bm{D}}
\newcommand{\J}{\bm{J}}
\newcommand{\Fceta}{\bm{F}_{c|\eta}}
\newcommand{\Fyeta}{\bm{F}_{y|\eta}}
\newcommand{\Fcalpha}{\bm{F}_{c|\alpha}}
\newcommand{\Fyalpha}{\bm{F}_{y|\alpha}}
\newcommand{\Fyalphatilde}{\widetilde{\bm F}_{y|\alpha}}
\newcommand{\FyalphaT}{\bm{F}_{y|\alpha,\theta}}
\newcommand{\R}{\bm{R}}
\newcommand{\V}{\bm{V}}
\newcommand{\U}{\bm{U}}
\newcommand{\I}{\bm{I}}
\newcommand{\HI}{\bm{H}_I}
\newcommand{\ee}{\mathrm{e}}
\newcommand{\order}{\mathcal{O}}
\newcommand{\deltamin}{\delta_{\mathrm{min}}}
\newcommand{\Mmax}{M_{\mathrm{max}}}
\newcommand{\B}{\bm{B}}
\newcommand{\rr}{\bm{r}}
\newcommand{\s}{\bm{s}}
\newcommand{\balpha}{\bm{\alpha}}
\newcommand{\bbeta}{\bm{\eta}}

\newcommand{\sigatom}{\sigma_{\mathrm{bead}}}

\newcommand{\SNR}{\mathrm{SNR}}
\newcommand{\signoise}{\sigma_{\mathrm{noise}}}
\newcommand{\sigsignal}{\sigma_{\mathrm{signal}}}

\DeclareMathOperator{\Tr}{Tr}
\DeclareMathOperator{\diag}{diag}
\DeclareMathOperator{\pdet}{pdet}
\DeclareMathOperator{\sech}{sech}
\DeclareMathOperator*{\argmax}{argmax}

\newcommand{\red}[1]{{\color{red} #1}}
\newcommand{\luke}[1]{{\color{blue}LE: #1}}
\newcommand{\henry}[1]{{\color{magenta}HM: #1}}


\begin{document}
	
	\title{Supporting Information for:\texorpdfstring{\\}{ }%
		Measurement-limited learning of conformational heterogeneity\texorpdfstring{\\}{ }%
        in cryo-electron microscopy}
	
	\author{Henry H. Mattingly}
	\email{hmattingly@flatironinstitute.org}
	\affiliation{Center for Computational Biology, Flatiron Institute, New York, NY, USA}
    
	\author{Luke Evans}
	\affiliation{Center for Computational Mathematics, Flatiron Institute, New York, NY, USA}
    
	\author{Pilar Cossio}
    \affiliation{Center for Computational Biology, Flatiron Institute, New York, NY, USA}
	\affiliation{Center for Computational Mathematics, Flatiron Institute, New York, NY, USA}

	
	\maketitle

    \begin{singlespace}
    \tableofcontents
    \end{singlespace}
\fi

	\section{Problem formulation}
    \label{SIsec:problem}
	
	\subsection{Discretizing a continuous latent space}
	
	Let $x \in \mathcal{X}$ denote a continuous latent variable describing the underlying physical state of a system (e.g., a molecular conformation). We seek to infer the probability distribution $P(x)$ over the latent variable from observed data $y \in \mathcal{Y}$, which is generated from $x$ through a noisy forward model,
	\begin{equation}
		P(y \mid x).
	\end{equation}
	A data set consists of $N$ IID observations $Y = \{y_1, \dots, y_N\}$.
	
	In many applications the latent space $\mathcal{X}$ is high-dimensional and continuous, making direct inference intractable. We therefore introduce a finite discretization of $\mathcal{X}$ by selecting $M$ representative structures
	\begin{equation}
		X = \{x_1,\dots,x_M\} \subset \mathcal{X}.
	\end{equation}
	The continuous latent variable can then be replaced by a categorical variable
	\begin{equation}
		z_n \in \{1,\dots,M\},
	\end{equation}
	where $z_n=m$ means observation $Y_n$ was generated from structure $X_m$, and $Z = (z_1, \dots, z_N)$.
	
	Thus,
	\begin{equation}
		P(x \mid \alpha) \approx \sum_{m=1}^M \alpha_m \, \delta(x-x_m),
	\end{equation}
	and the problem is then to infer the mixture weights, $\alpha$:
	\begin{equation}
		\alpha = (\alpha_1,\dots,\alpha_M), \qquad \alpha_m \ge 0, \quad \sum_{m=1}^M \alpha_m = 1,
	\end{equation}
	which represent the population fractions of the discretized states.
	
	We assume a Dirichlet prior on $\alpha$ \cite{bishop_pattern_2006, gelman_bayesian_2014},
	\begin{equation}
		\alpha \sim \Dir(\bm{\beta}),
	\end{equation}
	where we take the Dirichlet parameter vector $\bm{\beta} = (\beta_1, \dots, \beta_M) = \beta \, \ones$, where $\ones$ is the vector of all ones in $\mathbb{R}^M$, so that the prior is symmetric in $m$.
	
	The generative model for a data set of size $N$ is:
	\begin{align}
		\alpha &\sim \Dir(\beta \, \ones), \label{SIeq:generative_model_1}
		\\
		z_n \mid \alpha &\sim \Cat(\alpha), \quad n=1,\dots,N, \label{SIeq:generative_model_2}
		\\
		y_n \mid z_n=m &\sim Q_m(y_n), \label{SIeq:generative_model_3}
	\end{align}
	where $\Cat$ is the categorical distribution, and
	\begin{equation}
		Q_m(y) \equiv P(y \mid x_m) = P(y \mid z = m)
	\end{equation}
	is the single-observation likelihood induced by structure $X_m$.
	
	The resulting likelihood of a single observation $y$ is the mixture \cite{lindsay_mixture_1995}
	\begin{equation}
		P(y \mid \alpha) = \sum_{m=1}^M P(y \mid z = m) \, P(z = m \mid \alpha) = \sum_{m=1}^M Q_m(y) \, \alpha_m.
		\label{SIeq:single_likelihood}
	\end{equation}
	since $y$ is conditionally independent of $\alpha$ when given $z$. A data set of $N$ observations $Y$ has likelihood:
	\begin{equation}
		P(Y \mid \alpha) = \prod_{n=1}^N \sum_{m=1}^M Q_m(y_n) \, \alpha_m.
		\label{SIeq:mixture_likelihood}
	\end{equation}
	
	Our goal is to choose a discretization of the latent space $X = \{x_m\}_{m=1}^M$ that makes the data under the generative model, $Y$, maximally informative about the underlying population distribution, $\alpha$.

	\subsection{Sufficient statistics and natural parameters of the multinomial distribution}
	
	The core of this problem is to infer the parameters, $\alpha$, of a categorical distribution from corrupted observations, $Y$, of the underlying latent components, $Z$. A sufficient statistic for the parameters $\alpha$ from data $Z$ is the count vector $C$:
	\begin{equation}
		c_m = \sum_{n=1}^N \mathbf{1}\{z_n=m\}, \qquad C = (c_1,\dots,c_M),
	\end{equation}
	which quantifies how many data points were generated from structure $x_m$. For a data set of size $N$, we must have
	\begin{equation}
		\sum_{m=1}^M c_m = N.
	\end{equation}
	
	Since $C$ is a sufficient statistic for the parameters $\alpha$, $P(\alpha \mid C) = P(\alpha \mid Z)$. For example, with a Dirichlet prior, the posterior $P(\alpha \mid Z)$ is also Dirichlet, but it only depends on $C$ \cite{bishop_pattern_2006, gelman_bayesian_2014}:
	\begin{equation}
		\alpha \mid Z \sim \Dir(\beta \, \ones + C).
		\label{SIeq:alpha_given_Z}
	\end{equation}	
	
	With IID assignments $Z$ given $\alpha$, $C$ has a multinomial distribution
	\begin{equation}
		C \mid \alpha \sim \Mult(N, \alpha).
	\end{equation}	
	Thus, much of the problem structure will be inherited from the multinomial distribution. The multinomial distribution is a member of the exponential family, with sufficient statistic $C$ and natural parameters $\eta$ \cite{brown_fundamentals_1986, wainwright_graphical_2008}:
	\begin{equation}
		\eta = \log(\alpha) + \mathrm{constant} \in \mathbb{R}^M,
		\label{SIeq:eta_def}
	\end{equation}
	with reverse mapping
	\begin{equation}
		\alpha_m = \frac{\exp(\eta_m)}{\sum_k \exp(\eta_k)},
		\label{SIeq:alpha_of_eta}
	\end{equation}
    which demonstrates the invariance of $\alpha$ to adding the same constant to all entries of $\eta$.
	
	Lastly, we will repeatedly encounter the per-sample covariance of multinomial distribution, $\V(\alpha)$:
	\begin{equation}
		\V(\alpha) \equiv \D(\alpha) - \alpha \, \alpha^\top,
		\label{SIeq:V_def}
	\end{equation}
	where
	\begin{equation}
		\D(\alpha) \equiv \diag(\alpha)
		\label{SIeq:D_def}
	\end{equation}
	is a diagonal matrix with entries $\alpha$ along the diagonal, and $\Cov(C) = N \, \V(\alpha)$.

	\section{Expected information gain}
    \label{SIsec:MI_def}
	
	The mutual information between parameters $\alpha$ and the data $Y$ quantifies how much we learn about $\alpha$ from data \cite{shannon_mathematical_1948, lindley_measure_1956}:
	\begin{equation}
		\Iay = \mathbb{E}_{P(\alpha,Y)} \left[\log \frac{P(\alpha,Y)}{P(\alpha)P(Y)} \right].
	\end{equation}
	Equivalently,
	\begin{equation}
		\Iay = H(\alpha) - \mathbb{E}_{P(Y)} \left[ H(\alpha \mid Y) \right]
		\label{SIeq:MI_entropy_form}
	\end{equation}
	where $H(\alpha)$ is the entropy of the distribution of $\alpha$:
	\begin{equation}
		H(\alpha) \equiv - \mathbb{E}_{P(\alpha)} \left[ \log P(\alpha) \right],
	\end{equation}
	and
	\begin{equation}
		P(Y) = \int P(Y \mid \alpha) P(\alpha)\, d\alpha.
	\end{equation}
	Therefore, the quantity $I(\alpha;Y)$ measures the expected reduction in uncertainty about $\alpha$ after observing the data set $Y$. Maximizing this quantity over the choice of $\{x_m\}$ therefore yields a discretization that is most informative about the mixture weights. Since mutual information is invariant to invertible reparameterization, $\Iay = I(\eta; Y)$.

	\subsection{Information gain in terms of latent assignments}
    \label{SIsubsec:Iaz}
	
	The graphical structure of the model is
	\begin{equation}
		\alpha \rightarrow z_n \rightarrow y_n, \qquad n=1,\dots,N,
	\end{equation}
	with conditional independence across $n$ given $\alpha$. We can use this structure to rewrite the mutual information using the latent assignments $Z$. The chain rule of mutual information \cite{cover_elements_2006} states that for some random variables or vectors $X$, $Y$, $Z$, 
	\begin{equation}
		I(X; Y, Z) = I(X; Y \mid Z) + I(X;Z) = I(X; Z \mid Y) + I(X;Y),
	\end{equation}    
	where $I(X; Y,Z)$ is the mutual information between $X$ and the collection of random variables ${Y, Z}$,
	\begin{equation}
		I(X; Y,Z) = H(X) - \langle H(X \mid Y, Z)\rangle,
	\end{equation}
	and $I(X; Y \mid Z)$ is the conditional mutual information between $X$ and $Y$, given $Z$
	\begin{equation}
		I(X; Y \mid Z) = \langle H(X\mid Z)\rangle - \langle H(X \mid Y, Z)\rangle.
	\end{equation}
	
	Applying this chain rule to $I(\alpha; Z, Y)$ gives
	\begin{equation}
		I(\alpha; Z, Y) = I(\alpha; Y \mid Z) + \Iaz= I(\alpha; Z \mid Y) + \Iay.
	\end{equation}
	But due to the Markov structure, $\alpha$ is independent of the data $Y$ if the latent assignments $Z$ are known: $I(\alpha; Y \mid Z) = 0$. Thus, we have:
	\begin{equation}
		\Iay = \Iaz - I(\alpha ; Z \mid Y).
		\label{SIeq:MI_decomp}
	\end{equation}
	Since $C$ is a sufficient statistic for $Z$, we also have
	\begin{equation}
		\Iaz = \Iac, \qquad I(\alpha ; Z \mid Y) = \Iacy.
	\end{equation}

    The first term in Eq.~\eqref{SIeq:MI_decomp}, $\Iaz$, quantifies how informative the latent assignments would be about $\alpha$ if they were observed directly. This term depends only on the Dirichlet--categorical structure and is independent of the observation model. $\Iaz$ generally increases with $M$ when $N>M$, favoring models with more latent components.

    The second term, $I(\alpha;Z\mid Y)$, quantifies the information about $\alpha$ that remains in the latent assignments $Z$ after the observations $Y$ are known. If each observation $y_n$ uniquely identified its generating component $z_n$, then observing $Z$ would add no information beyond observing $Y$, so $I(\alpha;Z\mid Y)=0$ and $\Iay=\Iaz$. Conversely, overlap among the likelihoods $Q_m(y)$ creates ambiguity in the assignment from observations to latent components, making $I(\alpha;Z\mid Y)>0$. In the limit where $Y$ contains no information about $Z$, this penalty approaches $\Iaz$, and $\Iay$ vanishes.

    Thus, maximizing $\Iay$ requires balancing the potential information gained by adding latent components against the assignment ambiguity created when their observation distributions overlap.

	\section{Large-\texorpdfstring{$N$}{N} behavior of the mutual information}
    \label{SIsec:largeN_MI}

    \subsection{Gaussian approximation}
    \label{SIsubsec:Gaussian_approx}

    To study the behavior of $\Iay$ analytically, we use a large-$N$ Gaussian approximation. We represent $\alpha$ as a vector in $\mathbb{R}^M$, but it is constrained to the simplex $\sum_{m=1}^M \alpha_m = 1$. Thus, allowed variations of $\alpha$ lie in the $(M-1)$-dimensional tangent space orthogonal to $\ones$, the vector of all ones in $\mathbb{R}^M$. All covariance matrices below are therefore singular in the ambient $M$-dimensional representation, with nullspace spanned by $\ones$. 
    
    The per-sample Fisher information \cite{amari_methods_2000} will be a central object and is defined as
    \begin{equation}
    	\Fyalpha(\alpha)
    	=
    	-\E_{y|\alpha}\!\left[\nabla_\alpha^2 \log P(y\mid\alpha)\right]
    	=
    	\E_{y|\alpha}\!\left[
    	\nabla_\alpha \log P(y\mid\alpha)\,
    	(\nabla_\alpha \log P(y\mid\alpha))^\top
    	\right] \in \mathbb{R}^{M \times M}.
    	\label{SIeq:Fisher_alpha}
    \end{equation}
    Throughout this work, the parameter $\alpha$ appearing in the expectation $\E_{y|\alpha}[\, \cdot \, ]$ is the same parameter at which the likelihood is differentiated. Thus, $\Fyalpha(\alpha)$ quantifies the \textit{local} identifiability of the data-generating parameter value $\alpha$ from observations $y\sim P(y\mid\alpha)$. 

    Importantly, the Fisher information above treats the parameters $\alpha\in\mathbb{R}^M$ as independent. However, since $\alpha$ is restricted to the simplex, only parameter variations that lie in the tangent space of the simplex are allowed, $\ones^\top \dd\alpha=0$. Thus, while $\Fyalpha(\alpha)$ above is a useful representation for derivations below, only its projection onto the tangent space of the simplex is meaningful.

    For large $N$, the posterior is approximately Gaussian with covariance
    \begin{equation}
    	\Sigma_{\alpha|Y}^{-1}
    	\approx
    	N\,\Fyalpha(\alpha)+\Sigma_\alpha^{-1},
    	\qquad N\gg 1,
    	\label{SIeq:posterior_cov}
    \end{equation}
    where $\Sigma_\alpha$ is the covariance matrix of the Dirichlet prior. This is the usual large-sample Bernstein--von Mises form \cite{vaart_asymptotic_1998}, with the Fisher information evaluated at the data-generating parameters. For our purposes, these data-generating parameters is the same $\alpha$ over which $\Iay$ is averaged, so we use the same symbol throughout.
    
    Using the entropy form of the mutual information Eq.~\eqref{SIeq:MI_entropy_form} and approximating the prior and posterior entropies using Gaussian covariances, we obtain the standard large-sample Gaussian approximation to the mutual information \cite{clarke_information-theoretic_1990, clarke_jeffreys_1994}:
    \begin{equation}
	   \begin{aligned}
    		\Iay
		      &=
    		H(\alpha)-\avg{H(\alpha\mid Y)}_Y \\
	        &\approx
    		-\frac{1}{2}\,\E_\alpha\!\left[\log \pdet \Sigma_\alpha^{-1}\right]
		      +
		      \frac{1}{2}\,\E_Y\!\left[\log \pdet \Sigma_{\alpha|Y}^{-1}\right] \\
		      &\approx
		      \frac{1}{2}\,\E_\alpha\!\left[
		      \log \pdet \! \left( N\,\Fyalpha(\alpha)+\Sigma_\alpha^{-1} \right)
		      -
		      \log \pdet \Sigma_\alpha^{-1}
		      \right] \\
		      &=
		      \frac{1}{2}\,\E_\alpha\!\left[
		      \log \pdet \! \left(\I + N\,\Fyalpha(\alpha)\,\Sigma_\alpha\right)
		      \right].
	    \end{aligned}
	    \label{SIeq:Iay_Gaussian}
    \end{equation}
    The approximation in the third line replaces the average over datasets $Y$ by an average over the data-generating parameter $\alpha$, valid in the large-$N$ limit where posterior fluctuations concentrate near the data-generating value of $\alpha$. $\pdet$ indicates that the argument is projected onto the tangent space of simplex before computing the determinant of the resulting matrix in $\mathbb R^{M-1 \times M-1}$.

    For a symmetric Dirichlet prior with concentration parameter $\beta$,
    \begin{equation}
    	\Sigma_\alpha
    	=
    	\frac{1}{M(M\beta+1)}
    	\left(
    	\I-\frac{1}{M}\ones\ones^\top
    	\right).
    	\label{SIeq:dirichlet_cov}
    \end{equation}
    Projecting onto the simplex removes the $\ones \ones^\top$ term, so this matrix acts as a scalar multiple of the identity on the simplex tangent space. Then Eq.~\eqref{SIeq:Iay_Gaussian} reduces to a sum over the $M-1$ eigenvalues $\lambda_k(\alpha)$ of $\Fyalpha(\alpha)$ on tangent space of the simplex:
    \begin{equation}
    	\Iay
    	\approx
    	\frac{1}{2}\,\E_\alpha\!\left[
    	\sum_{k=1}^{M-1}
    	\log\!\left(
    	1+\frac{N\,\lambda_k(\alpha)}{M(M\beta+1)}
    	\right)
    	\right].
    	\label{SIeq:Iay_eigs}
    \end{equation}

    Equation \eqref{SIeq:Iay_eigs} shows that each identifiable parameter direction contributes one term to the mutual information. A direction contributes significantly when
    \begin{equation}
    	N\,\lambda_k(\alpha)\gg M(M\beta+1),
    \end{equation}
    and contributes negligibly when
    \begin{equation}
    	N\,\lambda_k(\alpha)\ll M(M\beta+1).
    \end{equation}
    Thus, increasing the number of mixture components can increase $\Iay$ only if the additional parameter directions remain identifiable.

    \subsection{Fisher information and the responsibility vector}
    \label{SIsubsec:Fisher_responsibilities}

Next, we derive the Fisher information matrix for the mixture-observation model. It is convenient to work first in terms of the natural parameters $\eta$ of the multinomial distribution, defined in Eq.~\eqref{SIeq:alpha_of_eta}, and then transform back to $\alpha$. The Fisher information with respect to $\eta$ is
\begin{equation}
	\Fyeta(\eta)
	\equiv
	\E_{y|\eta}\!\left[
	\nabla_\eta \log P(y\mid\eta)\,
	(\nabla_\eta \log P(y\mid\eta))^\top
	\right]
	=
	\E_{y|\eta}\!\left[s_y(\eta)\,s_y(\eta)^\top\right],
	\label{SIeq:Fisher_def}
\end{equation}
where
\begin{equation}
	s_y(\eta)\equiv \nabla_\eta \log P(y\mid\eta)
\end{equation}
is the score. Since the expectation of the score is zero, $\Fyeta(\eta)$ is also the covariance of $s_y(\eta)$. Furthermore, $\alpha(\eta)$ is invariant to adding a constant to all entries of $\eta$. Therefore, the likelihood $P(y \mid \eta)$ is unchanged along this direction, and $\Fyeta(\eta)$ has a null direction proportional to $\ones$.

Using the softmax parameterization
\begin{equation}
	\alpha_j(\eta)=\frac{e^{\eta_j}}{\sum_{\ell=1}^M e^{\eta_\ell}},
\end{equation}
we have
\begin{equation}
	\frac{\partial \alpha_j}{\partial \eta_m}
	=
	\alpha_j(\eta)\left(\delta_{jm}-\alpha_m(\eta)\right).
\end{equation}
Then, from the mixture likelihood Eq.~\eqref{SIeq:single_likelihood},
\begin{equation}
	P(y\mid\eta)=\sum_{j=1}^M Q_j(y)\,\alpha_j(\eta),
\end{equation}
the score is
\begin{equation}
\begin{aligned}
	(s_y(\eta))_m
	&=
	\frac{\partial}{\partial\eta_m}
	\log\!\left(\sum_{j=1}^M Q_j(y)\,\alpha_j(\eta)\right) \\
	&=
	\frac{1}{P(y\mid\eta)}
	\sum_{j=1}^M Q_j(y)\,
	\frac{\partial \alpha_j}{\partial \eta_m} \\
	&=
	\frac{1}{P(y\mid\eta)}
	\sum_{j=1}^M Q_j(y)\,
	\alpha_j(\eta)
	\left(\delta_{jm}-\alpha_m(\eta)\right) \\
	&=
	\frac{Q_m(y)\alpha_m(\eta)}{P(y\mid\eta)}
	-
	\alpha_m(\eta)
	\frac{\sum_{j=1}^M Q_j(y)\alpha_j(\eta)}{P(y\mid\eta)} \\
	&=
	\frac{Q_m(y)\alpha_m(\eta)}{\sum_{j=1}^M Q_j(y)\alpha_j(\eta)}
	-\alpha_m(\eta).
\end{aligned}
\end{equation}
This motivates the definition of the responsibility vector $r(y;\eta)\in\mathbb{R}^M$ with entries
\begin{equation}
	r_m(y;\eta)
	\equiv
	\frac{Q_m(y)\,\alpha_m(\eta)}{\sum_{j=1}^M Q_j(y)\,\alpha_j(\eta)}
	=
	P(z=m\mid y,\eta),
	\label{SIeq:r_def}
\end{equation}
that is, the conditional probability that latent component $m$ generated observation $y$. In terms of $r(y;\eta)$,
\begin{equation}
	s_y(\eta)=r(y;\eta)-\alpha(\eta).
	\label{SIeq:score_r_minus_alpha}
\end{equation}

Because $r(y;\eta)$ is a probability vector, $\sum_m r_m(y;\eta)=1$. Its expectation under the mixture likelihood is
\begin{equation}
	\E_{y|\eta}[r_m(y;\eta)]
	=
	\int
	\frac{Q_m(y)\,\alpha_m(\eta)}{P(y\mid\eta)}
	P(y\mid\eta)\,\dd y
	=
	\alpha_m(\eta).
\end{equation}

Substituting Eq.~\eqref{SIeq:score_r_minus_alpha} into Eq.~\eqref{SIeq:Fisher_def} gives
\begin{equation}
	\Fyeta(\eta)
	=
	\Cov_{y|\eta}(r(y;\eta))
	=
	\E_{y|\eta}\!\left[r(y;\eta)\,r(y;\eta)^\top\right]
	-
	\alpha(\eta)\,\alpha(\eta)^\top.
	\label{SIeq:Fisher_r_cov}
\end{equation}
Defining
\begin{equation}
	\R(\eta)\equiv \E_{y|\eta}\!\left[r(y;\eta)\,r(y;\eta)^\top\right],
\end{equation}
we have
\begin{equation}
	\Fyeta(\eta)=\R(\eta)-\alpha(\eta)\,\alpha(\eta)^\top.
\end{equation}

To connect this to overlap between the component image distributions, substitute Eq.~\eqref{SIeq:r_def} into $\R(\eta)$:
\begin{equation}
	\begin{aligned}
		(\R(\eta))_{m,m'}
		&=
		\E_{y|\eta}\!\left[r_m(y;\eta)\,r_{m'}(y;\eta)\right] \\
		&=
		\int
		\frac{Q_m(y)\,\alpha_m(\eta)}{P(y\mid\eta)}
		\frac{Q_{m'}(y)\,\alpha_{m'}(\eta)}{P(y\mid\eta)}
		P(y\mid\eta)\,\dd y \\
		&=
		\alpha_m(\eta)\,\alpha_{m'}(\eta)
		\int
		\frac{Q_m(y)\,Q_{m'}(y)}{P(y\mid\eta)}
		\,\dd y.
	\end{aligned}
\end{equation}
Therefore the entries of the Fisher information are
\begin{equation}
	(\Fyeta(\eta))_{m,m'}
	=
	\alpha_m(\eta)\,\alpha_{m'}(\eta)
	\left(
	\int
	\frac{Q_m(y)\,Q_{m'}(y)}{P(y\mid\eta)}
	\,\dd y
	-
	1
	\right).
	\label{SIeq:Fyeta_entries}
\end{equation}
Thus, $\Fyeta(\eta)$ is controlled by likelihood-weighted overlaps of the component image distributions $Q_m(y)$. When two components produce similar observations, their responsibilities are correlated across samples, reducing the identifiability of parameter directions that redistribute probability mass between them.

\subsection{Fisher information in \texorpdfstring{$\alpha$}{alpha} coordinates}
\label{SIsubsec:Fisher_alpha_coordinates}

We now derive the Fisher information whose eigenvalues enter the Gaussian mutual information approximation in Eq.~\eqref{SIeq:Iay_eigs}. The subtlety is that the mixture weights $\alpha$ lie on the simplex, so allowed perturbations satisfy $\ones^\top \dd\alpha=0$. We therefore first write the Fisher information as an $M\times M$ matrix in the ambient coordinates of $\alpha$, denoted $\Fyalpha(\alpha)$, and then restrict it to the $(M-1)$-dimensional tangent space of the simplex.

First, differentiate the likelihood with respect to the ambient coordinates of $\alpha$, treating the components of $\alpha$ as formally independent. From Eq.~\eqref{SIeq:single_likelihood},
\begin{equation}
    \frac{\partial}{\partial \alpha_m}\log P(y\mid\alpha)
    =
    \frac{Q_m(y)}{\sum_j Q_j(y)\alpha_j}
    =
    \frac{r_m(y;\alpha)}{\alpha_m}.
\end{equation}
This gives the ambient score vector
\begin{equation}
    g_\alpha(y;\alpha)
    =
    \D(\alpha)^{-1} \, r(y;\alpha).
\end{equation}
Because allowed perturbations satisfy $\ones^\top \dd\alpha=0$, adding any multiple of $\ones$ to this score does not change directional derivatives on the simplex. We therefore use the centered representative
\begin{equation}
    s_\alpha(y;\alpha)
    =
    \D(\alpha)^{-1} r(y;\alpha)-\ones,
    \label{SIeq:score_alpha_centered}
\end{equation}
which has zero mean under $P(y\mid\alpha)$. We define the ambient Fisher information in ambient $\alpha$ coordinates as
\begin{equation}
    \Fyalpha(\alpha)
    \equiv
    \E_{y|\alpha}\!\left[
    s_\alpha(y;\alpha) \, s_\alpha(y;\alpha)^\top
    \right].
\end{equation}
Using $\E_{y|\alpha}[r(y;\alpha)]=\alpha$, this becomes
\begin{equation}
    \Fyalpha(\alpha)
    =
    \D(\alpha)^{-1} \, \R(\alpha) \, \D(\alpha)^{-1}
    -
    \ones\ones^\top.
    \label{SIeq:F_alpha}
\end{equation}

The same matrix is obtained by transforming the natural-parameter Fisher information, $\Fyeta(\eta)$, derived in the previous subsection. Fixing the additive constant in
\begin{equation}
    \eta_m=\log\alpha_m+\mathrm{const},
\end{equation}
tangent perturbations obey
\begin{equation}
    \dd\eta=\D(\alpha)^{-1}\dd\alpha.
\end{equation}
Therefore Eq.~\eqref{SIeq:Fisher_r_cov} gives
\begin{equation}
    \Fyalpha(\alpha)
    =
    \D(\alpha)^{-1} \, \Fyeta(\eta(\alpha)) \, \D(\alpha)^{-1}
    =
    \D(\alpha)^{-1} \, \R(\alpha) \, \D(\alpha)^{-1}
    -
    \ones\ones^\top.
    \label{SIeq:F_alpha_alt}
\end{equation}
Although the same $-\ones\ones^\top$ term appears in Eqs.~\eqref{SIeq:F_alpha} and \eqref{SIeq:F_alpha_alt}, its interpretation is slightly different in the two derivations. In the direct $\alpha$-coordinate derivation, it comes from choosing a centered representative of the ambient score. In the change-of-variables derivation, it reflects the null direction of $\Fyeta(\eta)$ associated with the redundancy of the softmax parameterization: shifts $\eta\to\eta+c\ones$ leave $\alpha(\eta)$, and hence $P(y\mid\eta)$, unchanged.

Using the expression for $\R(\alpha)$ derived above, the entries of $\Fyalpha(\alpha)$ are
\begin{equation}
    \left(\Fyalpha(\alpha)\right)_{m,m'}
    =
    \E_{y|\alpha}\!\left[
    \frac{r_m(y;\alpha)}{\alpha_m}
    \frac{r_{m'}(y;\alpha)}{\alpha_{m'}}
    \right]
    -1
    =
    \int
    \frac{Q_m(y) \, Q_{m'}(y)}
    {\sum_j Q_j(y)\alpha_j}
    \,\dd y
    -1.
    \label{SIeq:Fyalpha_entries}
\end{equation}

Finally, to represent the Fisher information as an ordinary $(M-1)\times(M-1)$ matrix, we restrict $\Fyalpha(\alpha)$ to the simplex tangent space. Choose a matrix
\begin{equation}
    \B\in\mathbb R^{M\times(M-1)}
\end{equation}
whose columns form an orthonormal basis for this tangent space:
\begin{equation}
    \B^\top \B=\I_{M-1},
    \qquad
    \B^\top\ones=0.
    \label{SIeq:B_def}
\end{equation}
Any orthonormal basis for the null space of the vector $\ones$ will give equivalent results. Then, we define
\begin{equation}
    \Fyalphatilde(\alpha)
    \equiv
    \B^\top \Fyalpha(\alpha) \, \B.
    \label{SIeq:F_alpha_projected}
\end{equation}
Equivalently, if
\begin{equation}
    \tilde s_\alpha(y;\alpha)
    =
    \B^\top s_\alpha(y;\alpha),
    \label{SIeq:score_projected}
\end{equation}
then
\begin{equation}
    \Fyalphatilde(\alpha)
    =
    \E_{y|\alpha}\!\left[
    \tilde s_\alpha(y;\alpha) \, \tilde s_\alpha(y;\alpha)^\top
    \right].
\end{equation}
The eigenvalues of $\widetilde{\bm F}_{y|\alpha}(\alpha)$---the tangent-space eigenvalues of $\Fyalpha(\alpha)$---are the $M-1$ eigenvalues that enter Eq.~\eqref{SIeq:Iay_eigs}.

\subsection{Overlaps shrink the eigenvalues of the Fisher information}
\label{SIsubsec:overlaps_shrink_Fisher_eigs}

We now show that overlap between the component image distributions $Q_m(y)$ decreases the eigenvalues of the Fisher information matrix. This is closely related to the missing-information principle for latent-variable models \cite{louis_finding_1982}. In the no-overlap limit, observing $y$ identifies the latent component unambiguously, and the Fisher information from images reduces to the Fisher information of the multinomial counts model. With overlap, the image Fisher information is smaller in the PSD order, so all ordered eigenvalues decrease.

If the $Q_m(y)$ are non-overlapping, then each observation $y$ uniquely determines which component generated it. In this case, the responsibility vector is one-hot, so
\begin{equation}
	r(y;\eta)\in \{e_1,\dots,e_M\},
\end{equation}
where $e_m$ is the $m$th standard basis vector. Therefore
\begin{equation}
	r(y;\eta)\,r(y;\eta)^\top = \diag(r(y;\eta)),
\end{equation}
and taking the expectation over $y$ gives
\begin{equation}
	\R(\eta)
	=
	\E_{y|\eta}\!\left[r(y;\eta)\,r(y;\eta)^\top\right]
	=
	\E_{y|\eta}\!\left[\diag(r(y;\eta))\right]
	=
	\diag(\alpha(\eta))
	=
	\D(\alpha).
\end{equation}
Substituting into Eq.~\eqref{SIeq:Fisher_r_cov},
\begin{equation}
	\Fyeta(\eta)
	=
	\D(\alpha)-\alpha(\eta)\,\alpha(\eta)^\top
	=
	\V(\alpha),
\end{equation}
where $\V(\alpha)$ is the per-sample covariance matrix of the multinomial distribution from Eq.~\eqref{SIeq:V_def}. Thus, in the absence of overlap, the Fisher information from observing images is identical to that from observing latent counts directly.

To quantify the effect of overlap, write
\begin{equation}
	\Fyeta
	=
	\D-\alpha\alpha^\top-(\D-\R)
	=
	\V-\U,
	\label{SIeq:F_V_U}
\end{equation}
where
\begin{equation}
	\U
	\equiv
	\D-\R
	=
	\diag(\alpha)-\E_{y|\eta}[r\,r^\top]
	=
	\E_{y|\eta}\!\left[\diag(r)-r\,r^\top\right].
	\label{SIeq:U_def}
\end{equation}
Here, $\V(\alpha)$ is the Fisher information of the multinomial distribution in natural parameters, since for exponential families the Fisher information equals the covariance of the sufficient statistic \cite{brown_fundamentals_1986, wainwright_graphical_2008}.

We claim that $\diag(r)-r\,r^\top$ is positive semidefinite for every $r\in\Delta^{M-1}$. Indeed, for any $x\in\mathbb{R}^M$,
\begin{equation}
	\begin{aligned}
		x^\top \, (\diag(r) - r\,r^\top) \, x
		&=
		\sum_{m=1}^M r_m \, x_m^2
		-
		\left(\sum_{m=1}^M r_m \, x_m\right)^2 \\
		&=
		\Var_{m\sim r}(x_m)
		\ge 0.
	\end{aligned}
\end{equation}
The last expression is the variance of a randomly sampled element $x_m$ from the vector $x$, where the integer index $m$ is sampled with probability $r_m$. Therefore
\begin{equation}
	\diag(r)-r\,r^\top \succeq 0
\end{equation}
for all vectors $r$, and hence
\begin{equation}
	\U \succeq 0.
\end{equation}
From Eq.~\eqref{SIeq:F_V_U}, it follows that
\begin{equation}
	\Fyeta \preceq \V.
\end{equation}
Therefore every ordered eigenvalue of $\Fyeta$ is bounded above by the corresponding eigenvalue of $\V$:
\begin{equation}
	\lambda_k(\Fyeta)\le \lambda_k(\V), \qquad k=1,\dots,M-1.
\end{equation}
Equality holds in the no-overlap limit, where $r$ is one-hot, $\diag(r)-r\,r^\top=0$, and $\U=0$. Thus, $\U$ represents the reduction in Fisher information due to observing only $y$ rather than the latent assignment $z$.

This proves that overlap between the component image distributions shrinks the eigenvalues of the Fisher information. Intuitively, when several components produce similar images, redistributing mixture weight among them changes the observation distribution only weakly. The corresponding directions in parameter space are therefore poorly identifiable, producing small Fisher eigenvalues and, ultimately, reducing the mutual information between parameters and data.

\section{Optimal discretization in a 1D Gaussian model}
\label{SIsec:1D_model}

We now specialize to a one-dimensional model in which the effect of overlap among likelihoods can be studied analytically. Our goal is to quantify how overlap reduces Fisher information for the mixture weights and, through the large-$N$ approximation to $\Iay$, predict the optimal spacing $\Delta$---equivalently, the optimal number of components $M$---for discretizing the latent space. The resulting optimum is specific to this model, but the mechanism that produces it is general.

\subsection{Problem setup}

We consider an underlying one-dimensional latent variable
\begin{equation}
x \in [0,L] \subset \mathbb{R}.
\end{equation}
To approximate this continuous latent space, we choose a discrete set of $M$ equally spaced representative points,
\begin{equation}
x_m = (m-1)\,\Delta, \qquad m=1,\dots,M,
\end{equation}
with spacing
\begin{equation}
\Delta = \frac{L}{M-1}.
\label{SIeq:Delta_M_relation}
\end{equation}
More generally, the representative points need not lie on a grid, but the evenly spaced case is analytically tractable.

Observations $y \in \mathbb{R}$ are generated from each representative point through a Gaussian forward model,
\begin{equation}
Q_m(y) = \mathcal{N}(y; x_m, \sigma^2),
\end{equation}
where $\sigma$ is the measurement noise. Thus, conditional on component $m$, observations are centered at $x_m$ with variance $\sigma^2$.

To evaluate $\Iay$, we need the eigenvalues of the Fisher information appearing in Eq.~\eqref{SIeq:Iay_eigs}. In general, their dependence on component spacing cannot be solved analytically, even in this simple model. We therefore make simplifying assumptions that expose the structure of $\Fyeta(\eta)$ and allow us to derive an approximate optimal spacing $\Delta$.

\subsection{Taylor expansion of the mutual information}
\label{SIsubsec:1D_Taylor_exp}

To derive $\Iay$ analytically, we Taylor expand the argument of the expectation over $\alpha$ in Eq.~\eqref{SIeq:Iay_eigs} about the prior mean, $\bar \alpha = \ones / M$. First, we can write $\Iay$ as
\begin{align}
    \Iay &\approx \frac12 \E_\alpha\!\left[\log \det \!\left(\I_{M-1} + \frac{N}{M \, (M\, \beta + 1)}  \Fyalphatilde(\alpha) \right) \right]
    \\
    &= \frac12 \E_\alpha\!\left[\Tr \log \!\left(\I_{M-1} + \frac{N}{M \, (M\, \beta + 1)}  \Fyalphatilde(\alpha) \right) \right] ~
    \\
    &\equiv \E_\alpha\!\left[ I(\alpha) \right] ~,
\end{align}
where $\Fyalphatilde(\alpha) \in \mathbb{R}^{M-1 \times M-1}$ is the Fisher information projected onto a basis in the tangent space of the simplex (Eq.~\eqref{SIeq:F_alpha_projected}), and $\Tr$ is the trace. We expand around the prior mean, $\alpha = \bar \alpha + \delta\alpha = \bar \alpha + \B \, \delta\tilde\alpha$, where $\delta\tilde\alpha\in \mathbb{R}^{M-1}$ lies in the simplex tangent space and $\B \in \mathbb{R}^{M\times(M-1)}$ in Eq.~\eqref{SIeq:B_def} maps the tangent space to ambient $\alpha$ coordinates. Writing the perturbations $\delta \alpha$  in terms of $\B$ ensures that the perturbations about the mean remain in the simplex. Expanding gives: 
\begin{align}
    \Iay &\approx \E_{\delta\alpha}\!\left[ I(\bar\alpha) 
    + (\nabla_\alpha \, I(\bar\alpha))^\top  \,  \delta \alpha 
    + \frac12 \, \delta \alpha^\top \, \HI(\bar\alpha) \, \delta \alpha 
    + \cdots \right] ~
    \\
    &= \E_{\delta\tilde\alpha}\!\left[ I(\bar\alpha) 
    + (\nabla_\alpha \, I(\bar\alpha))^\top \,\B \,  \delta \tilde \alpha 
    + \frac12 \, \delta \tilde\alpha^\top \, \B^\top \HI(\bar\alpha) \, \B \, \delta \tilde\alpha 
    + \cdots \right] ~,
\end{align}
where $\HI(\bar\alpha)$ is the Hessian matrix of $I(\alpha)$ with respect to $\alpha\in\mathbb{R}^M$ evaluated at $\alpha = \bar\alpha$, and $\nabla_\alpha \, I(\bar\alpha)$ is the gradient evaluated at $\bar\alpha$ and is a column vector.

The expectation of the linear-order term is always zero when the expansion is about the mean, $\E[\delta\alpha]=\B \, \E[\delta\tilde\alpha]=\bm{0}_M$. Therefore, using the cyclic property of the trace, the leading order correction is quadratic in $\delta\alpha$:
\begin{equation}
    \Iay \approx I(\bar\alpha) + \frac12 \, \Tr[\B^\top \HI(\bar\alpha) \, \B \, \Sigma_{\delta\tilde\alpha}] + \cdots ~,
\end{equation}
where $\Sigma_{\delta\tilde\alpha} \in \mathbb{R}^{M-1 \times M-1}$,  with $\Sigma_{\delta\tilde\alpha} = \E[\delta\tilde\alpha  \, \delta\tilde\alpha^{\top}]$, is the covariance of the perturbations in the simplex tangent space under the Dirichlet prior. In particular,
\begin{align}
    \Sigma_{\delta\tilde\alpha} &= \E_\alpha\!\left[\B^\top(\alpha - \bar\alpha) \, (\alpha - \bar\alpha)^\top \B \right]
    \\
    &= \B^\top \, \Sigma_\alpha \, \B
    \\
    &= \frac{1}{M\, (M\,\beta+1)} \I_{M-1} ~,
\end{align}
where the last step used $B^\top \ones_M = \bm{0}_{M-1}$. Thus,
\begin{equation}
    \Iay \approx I(\bar\alpha) + \frac12 \, \frac{1}{M\, (M\,\beta+1)} \, \Tr[\B^\top \HI(\bar\alpha) \, \B] + \cdots ~.
    \label{SIeq:Iay_2nd_order}
\end{equation}

Finally, after we complete the derivation of $\Iay$ for this problem, we will show that the order-$n$ corrections in Eq.~\eqref{SIeq:Iay_2nd_order} scale as $\order(M^{-(n-1)})$ for large $M$. Furthermore, close to $M^*$, $I(\bar\alpha) \sim \order(M)$, so the corrections are at most a factor $\order(M^{-2})\ll1$ smaller than $I(\bar\alpha)$. Thus, we approximate $\Iay$ with
\begin{equation}
    \Iay \approx I(\bar\alpha) = \frac12 \sum_{k=1}^{M-1} \log\left(1 + \frac{N \, \lambda_k(\bar\alpha)}{M \, (M\,\beta+1)} \right),
    \label{SIeq:Iay_1D_leading}
\end{equation}
where $\lambda_k(\bar\alpha)$ are the eigenvalues of $\Fyalphatilde(\bar\alpha)$.

\subsection{Graph Laplacian structure of the Fisher information}
\label{SIsubsec:1D_graph_laplace}
Evaluated at $\bar\alpha = \ones/M$, the Fisher information matrix gains useful structure. First, using Eqs.~\eqref{SIeq:F_alpha_alt} and \eqref{SIeq:F_V_U}, we can generically write $\Fyalpha(\alpha)$ as
\begin{equation}
    \Fyalpha(\bar\alpha) = \D^{-1}(\alpha) \, \left( \V(\alpha) - \U(\alpha) \right) \, \D^{-1}(\alpha)~,
\end{equation}
where $\U(\alpha) = \D(\alpha) - \R(\alpha)$.

Importantly, $\U(\alpha)$ has the structure of a graph Laplacian. To show this, consider a graph with $M$ nodes, corresponding to the $M$ latent components, and symmetric edge weights 
\begin{equation}
w_{m,m'}(\alpha) \equiv \E_{y|\alpha}[ r_m(y;\alpha) \, r_{m'}(y;\alpha) ].
\label{SIeq:wmm'_def}
\end{equation}
Then the off-diagonal terms of $\U$ are 
\begin{equation}
(\U)_{m,m'}(\alpha) = -w_{m,m'}(\alpha), \quad m'\neq m.
\end{equation}
As required for a graph Laplacian, the all-ones vector $\ones$ is in the nullspace of $\U$:
\begin{equation}
\U(\alpha) \, \ones = \alpha - \E_{y|\alpha}[r(y;\alpha) \, r(y;\alpha)^\top \ones] = \alpha - \E_{y|\alpha}[r(y;\alpha)] = \alpha - \alpha = 0.
\end{equation}
This implies that each row of $\U$ sums to zero, and thus each diagonal element is the degree of its associated node:
\begin{equation}
\begin{gathered}
	(\U)_{mm} + \sum_{m'\neq m} (\U)_{m,m'} = (\U)_{mm} - \sum_{m'\neq m} w_{m,m'} = 0 
	\\
	\implies (\U)_{mm} = \sum_{m'\neq m} w_{m,m'} = \deg_m,
\end{gathered}
\end{equation}
where $\deg_m$ is the degree of node $m$. Thus, $\U(\alpha)$ is the Laplacian of the ``confusion" graph defined by Eq.~\eqref{SIeq:wmm'_def}.

Next, we specialize to $\alpha = \bar \alpha$ and assume $M\gg1$ ($L\gg \sigma$), so that most components $x_m$ are far from the boundaries of the domain. In this case, the graph and its weights are approximately translation-invariant. Then, the confusion weights $w_{m,m+j}$ are independent of $m$
\begin{equation}
w_{m,m+j} = \E_{y|\bar\alpha}[ r_m(y;\bar\alpha) \, r_{m+j}(y;\bar\alpha) ] \equiv w_j(\delta)
\label{SIeq:w_def}
\end{equation}
and only depend on the dimensionless spacing
\begin{equation}
\delta \equiv \Delta/\sigma.
\end{equation}
In the same regime, we neglect boundary effects and approximate the confusion graph $\U(\bar\alpha)$ as translation invariant and periodic. Thus, $\deg_m \approx d$ becomes independent of $m$, and $\U(\bar\alpha)$ is circulant,
\begin{equation}
\U(\bar\alpha) \approx
\begin{pmatrix}
	d & -w_{1} & -w_{2} & \cdots & -w_{M-1} \\
	-w_{1} & d & -w_{1} & \ddots & \vdots \\
	-w_2 & -w_{1} & d & \ddots & -w_2 \\
	\vdots & \ddots & \ddots & \ddots & -w_1 \\
	-w_{M-1} & \cdots & -w_2 & -w_1 & d
\end{pmatrix}.
\label{SIeq:U_circulant}
\end{equation}

Furthermore, at $\alpha = \bar\alpha$, $D(\bar\alpha) = \I/M$,
\begin{equation}
    \V(\bar\alpha) = \frac{1}{M} \left(\I - \frac{1}{M} \, \ones \ones^\top \right) ~,
\end{equation}
and 
\begin{equation}
    (\R(\bar\alpha))_{m,m'} = \E_{y|\bar\alpha} \! \left[r_m(y ; \bar\alpha) \, r_{m'}(y; \bar\alpha) \right] = \frac{1}{M^2} \int \frac{Q_m(y) \, Q_{m'}(y)}{P(y \mid \bar \alpha)} \, \dd y ~.
\end{equation}
With these, the Fisher information matrix projected onto the simplex tangent space in Eq.~\eqref{SIeq:F_alpha_projected} is 
\begin{equation}
    \Fyalphatilde(\bar\alpha) = M(\I_{M-1} - M \, \tilde\U(\bar\alpha)) ~,
    \label{SIeq:Fyalpha_bar_alpha}
\end{equation}
where we defined
\begin{equation}
    \tilde\U(\bar\alpha) = \B^T \, \U(\bar\alpha) \, \B~.
\end{equation}
Thus, by finding the eigenvalues of $\tilde\U(\bar\alpha)$ along directions within the simplex tangent space, we can compute the eigenvalues of $\Fyalphatilde(\bar\alpha)$ needed to evaluate the approximation of $\Iay$ in Eq.~\eqref{SIeq:Iay_1D_leading}. 

Since $\U(\bar\alpha)$ is approximately circulant, it is diagonalized by discrete Fourier modes:
\begin{equation}
\phi_k(m)=\exp\!\left(\frac{2\pi i \, k \, m}{M}\right),\qquad k=0,1,\dots,M-1.
\end{equation}
The $k=0$ mode is $\phi_0 = \ones$ and is perpendicular to the tangent space of the simplex. Thus, projecting on the simplex will eliminate this mode and its eigenvalue from the spectrum of $\tilde\U(\bar\alpha)$.

The eigenvalues corresponding to the remaining modes in the simplex tangent space are:
\begin{equation}
\lambda_k(\tilde \U(\bar\alpha)) \approx 2 \, \sum_{j\geq1} w_j(\delta) \left(1-\cos\!\left(\frac{j\, 2\,\pi \,k}{M}\right)\right) = 4 \, \sum_{j\geq1} w_j(\delta) \, \sin^2\!\left(\frac{j\, \pi \, k}{M}\right)~,
\label{SIeq:ring_laplacian_eigs}
\end{equation}
for $k=1,\dots,M-1$. Thus, the eigenvalues of $\Fyalphatilde(\bar\alpha)$ are
\begin{equation}
\lambda_k(\bar\alpha) 
\approx M \left( 1
- 4 \, M \, \sum_{j\geq1} w_j(\delta) \, \sin^2\!\left(\frac{j \, \pi \, k}{M}\right)\right),
\qquad k=1,\dots,M-1.
\label{SIeq:eigs_uniform_nn}
\end{equation}

\subsection{Overlap integrals}

To continue, we need to say something about the ``graph" edge weights or overlap integrals, $w_j(\delta)$. Recalling from Eq.~\eqref{SIeq:w_def}, these are defined as:
\begin{align}
w_j(\delta) &= \E_{y|\alpha}[ r_m(y;\alpha) \, r_{m+j}(y;\alpha) ] 
\\
&= \E_{y|\alpha}[ r_0(y;\alpha) \, r_{j}(y;\alpha) ]
\\
&=\frac{1}{M} \, \int_{-\infty}^{\infty} \frac{Q_0(y) \, Q_j(y)}{\sum_m Q_m(y)}
\end{align}
for large $M$ and far from the boundaries. 

The denominator has the form of a periodic sum Gaussian distributions spaced by $\Delta$. In terms of a Gaussian with standard deviation $\sigma$, $\phi_\sigma(t)$:
\begin{equation}
\sum_{m\in\mathbb{Z}} Q_m(y) = \sum_{m\in\mathbb{Z}} \phi_\sigma\!\left(y - \Delta \, m \right).
\label{SIeq:Qsum}
\end{equation}
Using Poisson summation, this can be written in terms of the Fourier transform of $\phi$, $\hat \phi_\sigma$:
\begin{equation}
\sum_{m\in\mathbb{Z}} \phi_\sigma\left(y - \Delta \, m \right) 
= \frac{1}{\Delta} \sum_{n\in\mathbb{Z}} \hat \phi_\sigma \! \left(\frac{2\pi\, n}{\Delta}\right) e^{i \, 2\pi \, n \, y/\Delta} 
= \frac{1}{\Delta} \sum_{n\in\mathbb{Z}} \exp\!\left(-\frac{2 \pi^2 \,n^2}{\delta^2} \right) e^{i \, 2\pi \, n \, y / \Delta}.
\end{equation}
The imaginary parts of $\pm n$ terms cancel out, leaving:
\begin{equation}
\sum_{m\in\mathbb{Z}} \phi_\sigma\left(y - \Delta \, m \right) 
= \frac{1}{\Delta}\left[1 + 2\, \sum_{n=1}^\infty \exp\!\left(-\frac{2 \pi^2 \,n^2}{\delta^2} \right) \cos\!\left( \frac{2\pi \, n \, y}{\Delta} \right) \right].
\end{equation}
For $\delta=O(1)$ or smaller, the nonzero Fourier modes are exponentially suppressed, so the Gaussian comb is approximately flat:
\begin{equation}
\sum_{m\in\mathbb Z} Q_m(y) \approx \frac{1}{\Delta}.
\label{SIeq:gauss_comb}
\end{equation}
Therefore
\begin{equation}
w_j(\delta)
\approx
\frac{\Delta}{M}
\int_{-\infty}^{\infty} Q_0(y) \, Q_j(y)\,\dd y.
\end{equation}
Since the product of two Gaussians can be integrated exactly,
\begin{equation}
\Delta\int_{-\infty}^{\infty} Q_0(y) \, Q_j(y)\,\dd y
=
\frac{\delta}{2\sqrt{\pi}}
\exp\!\left(-\frac{j^2\delta^2}{4}\right),
\end{equation}
we obtain
\begin{equation}
w_j(\delta)
\approx
\frac{1}{M}
\frac{\delta}{2\sqrt{\pi}}
\exp\!\left(-\frac{j^2\delta^2}{4}\right).
\label{SIeq:wj_approx}
\end{equation}

\subsection{Eigenvalues in closed form}
\label{SIsubsec:1D_eigenvalues}

We now simplify the Fisher eigenvalues in Eq.~\eqref{SIeq:eigs_uniform_nn} using the approximate overlap weights in Eq.~\eqref{SIeq:wj_approx}. The key point is that the weights are Gaussian in the component separation $j$, so their discrete Fourier transform can be evaluated using Jacobi theta functions. In the small-$\delta$ limit, this yields a simple expression for the Fisher spectrum.

Let
\begin{equation}
    \theta_k \equiv \frac{2\pi k}{M}.
\end{equation}
Substituting Eq.~\eqref{SIeq:wj_approx} into Eq.~\eqref{SIeq:eigs_uniform_nn} gives
\begin{equation}
\begin{aligned}
	\lambda_k(\bar\alpha)
	&\approx
	M \left[
	1-\frac{\delta}{\sqrt{\pi}}
	\sum_{j=1}^\infty
	e^{-j^2 \, \delta^2/4}
	\left(1-\cos(j \, \theta_k)\right)
	\right].
\end{aligned}
\label{SIeq:lambda_sum}
\end{equation}
The sums in Eq.~\eqref{SIeq:lambda_sum} are Jacobi theta functions. Using
\begin{equation}
\vartheta_3(z,q)
=
1+2\sum_{j=1}^{\infty}q^{j^2}\cos(2 \, j \, z),
\end{equation}
with $q=e^{-\delta^2/4}$, we obtain
\begin{equation}
\lambda_k(\bar\alpha)
\approx
M\left[
1-\frac{\delta}{2\sqrt{\pi}}
\left(
\vartheta_3(0,e^{-\delta^2/4})
-
\vartheta_3\!\left(\frac{\theta_k}{2},e^{-\delta^2/4}\right)
\right)
\right].
\label{SIeq:lambda_theta}
\end{equation}

To extract the small-$\delta$ behavior, we apply Poisson summation to the theta function series, rewriting it as a sum of Gaussians in Fourier space:
\begin{equation}
\vartheta_3(z,e^{-\delta^2/4})
=
\frac{2\sqrt{\pi}}{\delta}
\sum_{n\in\mathbb Z}
\exp\!\left[
-\frac{(z-\pi \,n)^2}{\delta^2/4}
\right].
\label{SIeq:theta_modular}
\end{equation}
For $z=\theta_k/2=\pi \, k/M \in [0,\pi]$ and $\delta\ll 1$, the dominant terms are $n=0$ and $n=1$:
\begin{equation}
\vartheta_3(z,e^{-\delta^2/4})
\approx
\frac{2\sqrt{\pi}}{\delta}
\left[
\exp\!\left(-\frac{z^2}{\delta^2/4}\right)
+
\exp\!\left(-\frac{(z-\pi)^2}{\delta^2/4}\right)
\right].
\end{equation}
Substituting this approximation into Eq.~\eqref{SIeq:lambda_theta} gives
\begin{equation}
\lambda_k(\alpha)
\approx
M\left[
\exp\!\left(-\frac{(2\pi \, k/M)^2}{\delta^2}\right)
+
\exp\!\left(-\frac{(2\pi(1-k/M))^2}{\delta^2}\right)
\right].
\label{SIeq:lambda_small_delta}
\end{equation}
Thus the smallest eigenvalue occurs at the highest-frequency mode, $k=M/2$:
\begin{equation}
\lambda_{\mathrm{min}}(\bar\alpha)
\approx
2 \, M \, e^{-\pi^2/\delta^2}.
\label{SIeq:lambda_min_small_delta}
\end{equation}
Thus, the eigenvalues of high-frequency parameter modes are exponentially suppressed when components are closely spaced compared to the observation width.

\subsection{Optimal spacing}
\label{SIsubsec:1D_optimal_delta}

We now optimize the approximate mutual information over the spacing $\delta=\Delta/\sigma$, or equivalently over $M$ at fixed $\Lambda=L/\sigma$. Using the symmetry $\lambda_k=\lambda_{M-k}$, the prefactor $1/2$ in Eq.~\eqref{SIeq:Iay_eigs} cancels for paired modes $k$ and $M-k$, leaving only a factor $1/2$ for the unpaired Nyquist mode $k=M/2$. Furthermore, in each eigenvalue except for the $M/2$ mode, one of the two exponentials in Eq.~\eqref{SIeq:lambda_small_delta} dominates. Using Eq.~\eqref{SIeq:lambda_small_delta},
\begin{equation}
\begin{aligned}
	\Iay &\approx \sum_{k=1}^{M/2-1} \log\!\left(1 + \frac{N}{M \, \beta + 1} e^{-4\pi^2 \, (k/M)^2/\delta^2} \right)
	\\
	&\qquad + \frac{1}{2}\log\!\left(1+\frac{2N}{M\, \beta+1}e^{-\pi^2/\delta^2}\right).
\end{aligned}
\end{equation}
For large $M$, $M\approx \Lambda/\delta$. Substituting this relation, the exponents of the bulk modes depend on $k/\Lambda$ rather than directly on $\delta$:
\begin{equation}
\Iay \approx
\sum_{k=1}^{M/2-1}
\log\!\left(1+\frac{N}{M \,\beta+1}e^{-(2\pi \, k/\Lambda)^2}\right)
+
\frac{1}{2}\log\!\left(1+\frac{2\, N}{M \, \beta+1}e^{-\pi^2/\delta^2}\right).
\end{equation}
Thus decreasing $\delta$ adds more modes, but it also reduces the identifiability of all modes through the factor $1/(M \, \beta+1)$ and exponentially suppresses the Nyquist mode.

Since adding terms is generally favorable as long as they are identifiable, we assume that near the optimal spacing only the $M/2$ mode is borderline identifiable. Mathematically, this along with taking $M$ large and $\beta\geq1$, gives:
\begin{equation}
\Iay \approx \sum_{k=1}^{M/2-1} \log\!\left(\frac{N}{M \, \beta} \, e^{-(2\pi\,k/\Lambda)^2} \right)
+ \frac{1}{2} \log\!\left(1 + \frac{2\, N}{M \, \beta} \, e^{-\pi^2\,M^2/\Lambda^2}\right).
\label{SIeq:Iay_bulk_id}
\end{equation}

Now suppose we add two structures, $M \rightarrow M + 2$ and $\delta = \Lambda/M \rightarrow \Lambda/(M+2)$. The mutual information $\Iay$ becomes:
\begin{equation}
\Iay^{+2} \approx \sum_{k=1}^{M/2} \log\!\left(\frac{N}{(M+2) \, \beta} \, e^{-(2\pi\,k/\Lambda)^2} \right)
+ \frac{1}{2} \log\!\left(1 + \frac{2\, N}{(M+2) \, \beta} \, e^{-\pi^2\,(M+2)^2/\Lambda^2}\right).
\end{equation}
Since $k$ itself does not depend on $M$, all but the $k=M/2$ term cancels when we take the difference of the two expressions above:
\begin{equation}
\begin{aligned}
	\Delta\Iay &= \frac{M}{2} \log\!\left(\frac{N}{(M+2) \, \beta}\right) - \left(\frac{M}{2}-1\right) \log\!\left(\frac{N}{M \, \beta} \right) - \left(\frac{\pi\,M}{\Lambda}\right)^2 + \dots
	\\
	&= \log\!\left(\frac{N}{(M+2) \, \beta}\right) - \left(\frac{M}{2}-1\right) \log\!\left(1+\frac{2}{M} \right) - \left(\frac{\pi}{\delta}\right)^2 + \dots
	\\
	&\approx \log\!\left(\frac{N}{M \, \beta}\right) - 1 - \left(\frac{\pi}{\delta}\right)^2,
\end{aligned}
\end{equation}
where we will neglect the small change in the $M/2$ eigenvalue, and in line 3 we take $M$ large to expand the $\log$.

At the optimum, adding a structure (or two) should cause roughly zero change in the mutual information. This gives us the following optimality condition:
\begin{equation}
\Delta\Iay = 0 \approx \log\!\left(\frac{N \, \delta^*}{\Lambda \, \beta}\right) - 1 - \left(\frac{\pi}{\delta^*}\right)^2.
\end{equation}
Thus, the optimal spacing is roughly set by balancing the gain of adding a new identifiable mode against the cost of reducing the identifiability of all existing modes. We can solve this in terms of the Lambert W function $x = W(y)$, which solves $x\,e^x = y$:
\begin{equation}
\begin{gathered}
	\frac{N \, \delta^*}{\Lambda \, \beta} \, e^{-\pi^2/(\delta^*)^2} = e 
	\\
	\frac{2 \, N^2 \, \pi^2}{\Lambda^2 \, \beta^2 \, e^2} = \frac{2 \, \pi^2}{(\delta^*)^2} \, e^{2\, \pi^2/(\delta^*)^2},
	\\
	\frac{2 \, \pi^2}{(\delta^*)^2} = W\!\left(\frac{2 \, \pi^2 \, N^2}{\Lambda^2 \, \beta^2 \, e^2}\right),
	\\
	\delta^* \approx \sqrt{2} \, \pi \, W\!\left(\frac{2 \, \pi^2 \, N^2}{\Lambda^2 \, \beta^2 \, e^2}\right)^{-1/2}
\end{gathered}
\end{equation}
For large $x$, $W(x) \approx \log(x) - \log \log (x)$, so for large $N$ the optimal spacing scales as:
\begin{equation}
\delta^* \approx \frac{\pi}{\sqrt{\log\!\left(\frac{\sqrt{2} \, \pi \, N}{\Lambda \, \beta \, e}\right)}},
\end{equation}
and therefore $\delta^*$ goes to zero \textit{very} slow with $N$, $\delta^* \sim (\log N)^{-1/2}$. Furthermore, the scale of the optimal spacing is set by the measurement noise $\sigma$.

At this optimal spacing, the least identifiable mode is marginally resolved in the mutual-information expression:
\begin{equation}
1+\frac{N \, \lambda_{\min}(\bar\alpha)}{M \, (M\, \beta+1)}
\approx
1+\frac{2 \, N}{M \, \beta}e^{-\pi^2/(\delta^*)^2}
\approx
1+2 \, e
\sim \order(1).
\end{equation}

Lastly, we use Eq.~\eqref{SIeq:Iay_bulk_id} to get an approximate closed form expression for $\Iay$ when $\delta$ is near the optimal value and $M^*$ is large:
\begin{equation}
\begin{aligned}
	\Iay &\approx \sum_{k=1}^{M^*/2-1} \log\!\left(\frac{N}{M^* \, \beta} \, e^{-(2\pi\,k/\Lambda)^2} \right)
	+ \frac{1}{2} \log\!\left(1 + 2\, \frac{N}{M^* \, \beta} \, e^{-\pi^2\,(M^*)^2/\Lambda^2}\right)
	\\
	&\quad
	= \left(\frac{M^*}{2}-1\right) \log\!\left(\frac{N}{M^* \, \beta}\right) - \left(\frac{2\pi}{\Lambda}\right)^2 \sum_{k=1}^{M^*/2-1} k^2
	+ \frac{1}{2} \log\!\left(1 + \frac{2\, N}{M^* \, \beta} \, e^{-\pi^2\,(M^*)^2/\Lambda^2}\right)
	\\
	&\quad
	\approx \left(\frac{M^*}{2}-1\right) \log\!\left(\frac{N}{M^* \, \beta}\right) - \frac{1}{3} \left(\frac{2\pi}{\Lambda}\right)^2 \left(\frac{M^*}{2}-1\right)^3
	+ \frac{1}{2} \log\!\left(1 + \frac{2\, N}{M^* \, \beta} \, e^{-\pi^2\,(M^*)^2/\Lambda^2}\right)
	\\
	&\quad
	\approx \frac{M^*-2}{2} \, \log\!\left(\frac{N}{M^* \, \beta}\right) + \frac{M^*-2}{2} \, \log\left(\exp\left(-\frac{1}{3} \left(\frac{\pi \, M^*}{\Lambda}\right)^2 \right)\right)
	+ \frac{1}{2} \log\!\left(1 + \frac{2\, N}{M^* \, \beta} \, e^{-\pi^2\,(M^*)^2/\Lambda^2}\right)
	\\
	&\quad
	\approx \frac{M^*-2}{2} \, \log\!\left(\frac{N}{M^* \, \beta} \, e^{-\pi^2 / 3\, \delta^2}\right)
	+ \frac{1}{2} \log\!\left(1 + \frac{2\, N}{M^* \, \beta} \, e^{-\pi^2/\delta^2}\right),
\end{aligned}
\label{SIeq:Iay_closed}
\end{equation}
where we approximated $\sum_{k=1}^n k^2 \approx n^3/3$ in the third line. This form has the interpretation that $M-2$ paired parameter directions are well resolved, with effective Fisher eigenvalues of order $M\exp[-\pi^2/(3 \, \delta^2)]$, while the remaining Nyquist direction is marginally identifiable.

Plugging in
\begin{equation}
    M^* \approx L/\Delta^* \approx \frac{L}{\sigma \,\pi} \, \sqrt{\log\!\left(\frac{\sqrt{2} \, \pi \, N}{\Lambda \, \beta \, e}\right)},
\end{equation}
taking $M^*$ large, using $\delta \approx \Lambda/M$ with $\Lambda = L/\sigma$, and neglecting the last term associated with the unidentifiable direction, the leading order scaling of the optimal mutual information is
\begin{equation}
    \begin{aligned}
        \Iay &\approx \frac{1}{2} M^* \log\!\left(\frac{N}{M^* \, \beta}\right) - \frac{\pi^2}{6}  \frac{\Lambda^2}{(M^*)^2} M^* 
        \\
        &\approx \frac{1}{2} M^* \left(\log(N) -\log(M^* \, \beta)\right)
        \\
        &\approx \frac{1}{2} \frac{\Lambda}{\pi} \, \sqrt{\log\!\left(\frac{\sqrt{2} \, \pi \, N}{\Lambda \, \beta \, e}\right)} \log(N)
        \\
        &\approx \frac{1}{2} \frac{\Lambda}{\pi} (\log(N))^{3/2}~.
    \end{aligned}
\end{equation}

\subsection{Higher-order corrections to \texorpdfstring{$\Iay$}{I(alpha;Y)}}
\label{SIsubsec:1D_Taylor_correction}

We now return to the Taylor expansion in Eq.~\eqref{SIeq:Iay_2nd_order} and estimate the size of the higher-order corrections. Recall that
\begin{equation}
    \Iay
    =
    \E_\alpha[I(\alpha)]
    \approx
    I(\bar\alpha)
    +
    \frac{1}{2M(M\beta+1)}
    \Tr\!\left[
    \B^\top \HI(\bar\alpha)\B
    \right]
    +\cdots ,
    \label{SIeq:Iay_Taylor_recall}
\end{equation}
where $\bar\alpha=\ones/M$ and
\begin{equation}
    I(\alpha)
    =
    \frac12
    \Tr\log\!\left[
    \I_{M-1}
    +
    \frac{N}{M(M\beta+1)}
    \Fyalphatilde(\alpha)
    \right].
    \label{SIeq:g_alpha_revisit}
\end{equation}
Thus, the second-order correction is controlled by the curvature of $I(\alpha)$ at $\bar\alpha$ along simplex directions, $\B^\top \HI(\bar\alpha)\B$. We'll first address the scaling of this term, and then use similar reasoning to get at the scaling of all higher terms.

We first derive the curvature, $\HI(\alpha)$. Differentiating Eq.~\eqref{SIeq:g_alpha_revisit} gives
\begin{equation}
    \frac{\partial I}{\partial\alpha_i}
    =
    \frac12
    \frac{N}{M(M\beta+1)}
    \Tr\!\left[
    \left(
    \I_{M-1}
    +
    \frac{N}{M(M\beta+1)}
    \Fyalphatilde
    \right)^{-1}
    \frac{\partial \Fyalphatilde}{\partial\alpha_i}
    \right],
    \label{SIeq:dg_dalpha_revisit}
\end{equation}
and
\begin{align}
    (\HI(\alpha))_{i,j} = \frac{\partial^2 I}{\partial\alpha_i\partial\alpha_j}
    &=
    \frac12
    \frac{N}{M(M\beta+1)}
    \Tr\!\left[
    \left(
    \I_{M-1}
    +
    \frac{N}{M(M\beta+1)}
    \Fyalphatilde
    \right)^{-1}
    \frac{\partial^2 \Fyalphatilde}
    {\partial\alpha_i\partial\alpha_j}
    \right]
    \notag \\
    &\quad
    -
    \frac12
    \left(
    \frac{N}{M(M\beta+1)}
    \right)^2
    \Tr\!\left[
    \left(
    \I_{M-1}
    +
    \frac{N}{M(M\beta+1)}
    \Fyalphatilde
    \right)^{-1}
    \frac{\partial \Fyalphatilde}{\partial\alpha_j}
    \right.
    \notag \\
    &\hspace{4.2cm}
    \left.
    \times
    \left(
    \I_{M-1}
    +
    \frac{N}{M(M\beta+1)}
    \Fyalphatilde
    \right)^{-1}
    \frac{\partial \Fyalphatilde}{\partial\alpha_i}
    \right] .
    \label{SIeq:Hessian_g_matrix_form}
\end{align}

Next, we derive derivatives of the Fisher information appearing above. Using Eq.~\eqref{SIeq:F_alpha_alt}, the ambient Fisher information can be written as
\begin{equation}
    \left(\Fyalpha(\alpha)\right)_{m,m'}
    =
    \int
    \frac{Q_m(y)\,Q_{m'}(y)}
    {P(y\mid\alpha)}
    \dd y
    -1 .
    \label{SIeq:F_alpha_mixture_revisit}
\end{equation}
For $\delta=\Delta/\sigma\lesssim1$ and away from boundaries, the Gaussian comb derived in Eq.~\eqref{SIeq:gauss_comb} gives
\begin{equation}
    P(y\mid\bar\alpha)
    =
    \frac1M
    \sum_{m=1}^M Q_m(y)
    \approx
    \frac{1}{M\Delta}
    \approx
    \frac{1}{L}.
    \label{SIeq:Py_uniform_flat}
\end{equation}
Therefore, at $\alpha=\bar\alpha$,
\begin{equation}
    \frac{\partial}{\partial\alpha_i}
    \left(\Fyalpha\right)_{m,m'}
    =
    -
    \int
    \frac{Q_m(y)\,Q_{m'}(y)\,Q_i(y)}
    {P(y\mid\bar\alpha)^2}
    \, \dd y
    \approx
    -L^2
    \int
    Q_m(y)\,Q_{m'}(y)\,Q_i(y)
     \, \dd y ,
    \label{SIeq:dF_local_revisit}
\end{equation}
and
\begin{equation}
    \frac{\partial^2}{\partial\alpha_i\partial\alpha_j}
    \left(\Fyalpha\right)_{m,m'}
    =
    2
    \int
    \frac{Q_m(y)\,Q_{m'}(y)\,Q_i(y)\,Q_j(y)}
    {P(y\mid\bar\alpha)^3}
    \, \dd y
    \approx
    2L^3
    \int
    Q_m(y)\,Q_{m'}(y)\,Q_i(y)\,Q_j(y)
    \, \dd y .
    \label{SIeq:d2F_local_revisit}
\end{equation}

Equations~\eqref{SIeq:dF_local_revisit} and \eqref{SIeq:d2F_local_revisit} show that the Fisher derivatives are local and translation-covariant away from boundary indices. At $\bar\alpha$, we showed earlier that $\Fyalphatilde(\bar\alpha)$ is also approximately translation invariant (Eq.~\eqref{SIeq:Fyalpha_bar_alpha}), and therefore so is the prefactor
\begin{equation}
    \left(
    \I_{M-1}
    +
    \frac{N}{M(M\beta+1)}
    \Fyalphatilde(\bar\alpha)
    \right)^{-1}.
\end{equation}
Consequently, the full Hessian in Eq.~\eqref{SIeq:Hessian_g_matrix_form} is approximately translation invariant in the ambient indices:
\begin{equation}
    (\HI(\bar\alpha))_{ij}
    \approx
    h(|i-j|),
\end{equation}
away from boundaries.

The simplex projection removes the all-ones direction. Using $\B\B^\top=\I_M-\ones\ones^\top/M$,
\begin{equation}
    \Tr[\B^\top \HI(\bar\alpha)\B]
    =
    \Tr[\HI(\bar\alpha)]
    -
    \frac{1}{M}\ones^\top \HI(\bar\alpha)\ones .
\end{equation}
For an approximately circulant Hessian, this becomes
\begin{equation}
    \Tr[\B^\top \HI(\bar\alpha)\B]
    \approx
    M h(0)-\sum_{m=0}^{M-1} h(m).
\end{equation}
Because the kernels in Eqs.~\eqref{SIeq:dF_local_revisit} and \eqref{SIeq:d2F_local_revisit} are localized on the scale of the measurement noise, $h(m)$ decays with the separation $m$ in component index. Therefore $\sum_m h(m)=\order(1)$ and
\begin{equation}
    I^{(2)} \equiv \frac{1}{2M(M\beta+1)}\Tr[\B^\top \HI(\bar\alpha)\B]
    =
    \order(M^{-1})
\end{equation}
for $\beta = 1$.

Higher-order terms, $I^{(n)}$, have the form
\begin{equation}
    I^{(n)} \equiv \frac{1}{n!} \sum_{i_1, \cdots, i_n}
    \frac{\partial^n \, I(\alpha)}{\partial\alpha_{i_1} \cdots 
    \partial\alpha_{i_n} }
    \E\! \left[ 
    \delta\alpha_{i_1}
    \cdots
    \delta\alpha_{i_n}
    \right] 
    \Bigg|_{\alpha=\bar\alpha} ~,
    \label{SIeq:higher_terms}
\end{equation}
with $i_m\in[1, \dots, M]$. Since each $\delta\alpha_m$ is a deviation from the mean, $\bar\alpha$, the expectations above are centered moments of the symmetric Dirichlet distribution, which scale as at most 
\begin{equation}
    \E\! \left[ 
    \delta\alpha_{i_1}
    \cdots
    \delta\alpha_{i_n}
    \right] = \order(M^{-n}) ~.
    \label{SIeq:Dir_centered_moments}
\end{equation}
Furthermore, by similar arguments as above, the higher derivatives of $I(\alpha)$ are all local in component indices. Therefore, even as the number of components $M$ increases, the higher-order derivative tensor still only contains $\order(M)$ identical local perturbations and depends only on differences in component indices, away from boundaries. As a result, by similar reasoning as in the second-order term, sum over derivative indices in Eq.~\eqref{SIeq:higher_terms} is $\order(M)$. All together, this means all higher order terms scale as
\begin{equation}
    I^{(n)} = \order(M^{(1-n)})
    \label{SIeq:higher_scaling}
\end{equation}
for $n \geq 2$ and $\beta = 1$.

Finally, we compare this correction to the leading term. From Eq.~\eqref{SIeq:Iay_1D_leading},
\begin{equation}
	I(\bar\alpha)
	=
	\frac12
	\sum_{k=1}^{M-1}
	\log\!\left(
	1+
	\frac{N\lambda_k(\bar\alpha)}
	{M(M\beta+1)}
	\right).
\end{equation}
Near the optimal spacing at large $N$, a finite fraction of the $M-1$ Fisher modes contribute to this sum. Thus, $I(\bar\alpha)\sim \order(M)$ for $M$ in the vicinity of $M^*$ (as in Eq.~\eqref{SIeq:Iay_closed}). When $\delta\gg\delta^*$, all parameters are identifiable, so we still have $\I(\bar\alpha)=\order(M)$. When $\delta\ll\delta^*$, at worst the number of identifiable parameters saturates, so the sum in $I(\bar\alpha)$ above stops growing with $M$ and remains $\order(1)$.

Combining the scaling above with Eq.~\eqref{SIeq:higher_scaling}, the relative correction scales as
\begin{equation}
	\frac{
		\E[I(\alpha)] - I(\bar\alpha) 
	}{
		I(\bar\alpha)
	}
	= 
    \frac{\sum_{n=2}^\infty I^{(n)}}{I(\bar\alpha) }
    =
	\order(M^{-2}) ~,
	\label{SIeq:Taylor_relative_correction}
\end{equation}
and thus is negligible for large $M$. When $\delta\ll\delta^*$, if $I(\bar\alpha)=\order(1)$, the correction may scale scaling as $\order(M^{-1})$, which is still small for large $M$.

This 1D result used the following assumptions: 1) the components are evenly-spaced with spacing $\Delta$; 2) the domain is large enough that there are many components, so $M$ is large; and 3) the components are not non-overlapping ($\delta\lesssim 1$).

\section{Conditional mutual information with known image parameters}
\label{SIsec:conditional_MI}

In many experimental settings, each observation $y_i$ is acquired under conditions
that are partially known.
In cryo-EM, for example, the defocus $\Delta f_i$ is estimated from every micrograph
and is therefore treated as a known quantity for each image.
More generally, let $\theta_i$ denote the known parameters for image $i$, and
$\Theta=(\theta_1,\dots,\theta_N)$ the full collection across the dataset.
Because $\Theta$ is observed, the information that $Y$ carries about $\alpha$ should
be quantified by the \emph{conditional} mutual information,
\begin{equation}
    \IayT
    =
    H(\alpha) - \E_{Y,\Theta}\!\left[H(\alpha\mid Y,\Theta)\right],
    \label{SIeq:IayT_def}
\end{equation}
where $H(\alpha\mid\Theta)=H(\alpha)$ because $\alpha$ and $\Theta$ are independent
\textit{a priori}.

\subsection{Effect on the posterior covariance}

With known $\theta_i$, the single-observation likelihood becomes
$P(y_i\mid\alpha,\theta_i) = \sum_m \alpha_m\,Q_m(y_i;\theta_i)$, and the
corresponding per-image Fisher information is
\begin{equation}
    \FyalphaT(\alpha,\theta)
    =
    \E_{y\mid\alpha,\theta}\!\left[
        s(y;\alpha,\theta)\,s(y;\alpha,\theta)^\top
    \right],
    \label{SIeq:Fisher_perimage}
\end{equation}
where $s(y;\alpha,\theta)=\nabla_\alpha\log P(y\mid\alpha,\theta)$.
Because images are conditionally independent given $(\alpha,\theta)$, the posterior
precision generalizes from Eq.~\eqref{SIeq:posterior_cov} to a sum over images,
\begin{equation}
    \Sigma_{\alpha\mid Y,\Theta}^{-1}
    \approx
    \sum_{i=1}^{N} \FyalphaT(\alpha,\theta_i)
    + \Sigma_\alpha^{-1}.
    \label{SIeq:posterior_cov_T}
\end{equation}
For large $N$, the sample average of per-image Fisher informations concentrates around
its expectation over the marginal distribution $P(\theta)$,
\begin{equation}
    \frac{1}{N}\sum_{i=1}^{N}\FyalphaT(\alpha,\theta_i)
    \;\xrightarrow{N\to\infty}\;
    \bar{\bm{F}}_{y\mid\alpha}(\alpha)
    \;\equiv\;
    \E_{\theta}\!\left[\FyalphaT(\alpha,\theta)\right].
    \label{SIeq:Fisher_avg}
\end{equation}
Thus, the precision self-averages to
\begin{equation}
    \Sigma_{\alpha\mid Y,\Theta}^{-1}
    \approx
    N\,\bar{\bm{F}}_{y\mid\alpha}(\alpha) + \Sigma_\alpha^{-1},
    \qquad N\gg 1,
    \label{SIeq:posterior_cov_avg}
\end{equation}
which has the same form as Eq.~\eqref{SIeq:posterior_cov} with the single-image
Fisher replaced by the $\theta$-averaged Fisher $\bar{\bm{F}}_{y\mid\alpha}(\alpha)$.

\subsection{Gaussian approximation}

Substituting Eq.~\eqref{SIeq:posterior_cov_avg} into Eq.~\eqref{SIeq:Iay_Gaussian}
gives
\begin{equation}
    \IayT
    \approx
    \frac{1}{2}\,\E_\alpha\!\left[
        \log\pdet\!\left(
            \I + N\,\bar{\bm{F}}_{y\mid\alpha}(\alpha)\,\Sigma_\alpha
        \right)
    \right].
    \label{SIeq:IayT_Gaussian}
\end{equation}
For a symmetric Dirichlet prior, $\Sigma_\alpha$ again acts as a scalar multiple of the
identity on the simplex tangent space (Eq.~\eqref{SIeq:dirichlet_cov}), so this
reduces to
\begin{equation}
    \IayT
    \approx
    \frac{1}{2}\,\E_\alpha\!\left[
        \sum_{k=1}^{M-1}
        \log\!\left(
            1 + \frac{N\,\bar\lambda_k(\alpha)}{M(M\beta+1)}
        \right)
    \right],
    \label{SIeq:IayT_eigs}
\end{equation}
where $\bar\lambda_k(\alpha)$ are the eigenvalues of $\bar{\bm{F}}_{y\mid\alpha}(\alpha)$
on the simplex tangent space.
This is the analogue of Eq.~\eqref{SIeq:Iay_eigs}: the
structure and interpretation are identical, with the Fisher eigenvalues now those of
the $\theta$-averaged Fisher information.

\section{Computational methods}
\label{SIsec:methods}

\subsection{Markov Chain Monte Carlo}
\label{SIsec:MCMC}

A custom Markov Chain Monte Carlo (MCMC) code was used to sample posterior distributions of parameters in the 1D model. For varying ensembles, the same data $Y$ was used.

MCMC was initialized at the prior mean, $\alpha_0 = \frac{1}{M} \, \ones$, and the proposal distribution was Dirichlet:
\begin{equation}
	\alpha' \sim \Dir(a \, \alpha_n + b),
\end{equation}
for some scalar $a$ and $b = 0.1$, where $\alpha_n$ are the current parameters of the Markov chain and $\alpha'$ are the proposed parameters for the next step. $b$ was a small baseline parameter, and $a$ controlled how tight the proposal distribution was around the current parameters.

MCMC was performed in two phases. The first phase was used to determine $a$. Starting from an initial value of $a$, 500 proposal attempts were made and the acceptance fraction $f$ computed. If $f < 0.05$, $a \rightarrow 1.5 \, a$; if $f > 0.15$, $a \rightarrow a/1.5$. This was repeated until $|a - 0.1| < 0.05$. 

In the second phase, MCMC was run with fixed $a$ and $b$ until a prescribed number $N_\mathrm{MC}$ of proposals were accepted. We set $N_\mathrm{MC} = 2000 \, M + 1000$. Then the first 1000 accepted parameters were thrown out to remove the initial transient. 

\subsection{Molecular simulation data}
\label{SIsec:mol_sim_data}

We constructed ensembles from molecular dynamics (MD) simulations of of the RNA P4-P6 domain of the \textit{Tetrahymena thermophila} group I intron ribozyme \cite{kruger_self-splicing_1982, woodson_folding_2002, cech_ribozymes_2002, bonilla_cryo-em_2022}. Coarse-grained MD simulations were performed using the Martini2 force field \cite{uusitalo_martini_2017}, fixing the secondary structure elements and breaking tertiary structure contacts to enable the system to explore the open-to-closed conformational landscape. MD simulations were performed using GROMACS \cite{abraham_gromacs_2015}, starting from the closed crystal structure \cite{cate_crystal_1996} and using a time step of 0.005 ps. This model of the RNA contains $N_{\mathrm{beads}} = 1,036$ coarse grained beads, and multiple simulations were combined for a total of 8890 frames.

\subsection{Structural alignment}
\label{SIsec:struct_align}

For each conformation, we subtracted the centroid from all atom coordinates. We then aligned all conformations to a reference structure by rigid-body rotation, choosing the reference as the structure with minimum average RMSD to all other conformations. Alignments were computed using the Kabsch algorithm \cite{kabsch_solution_1976}. In rare cases where the Kabsch alignment failed numerically, we minimized the RMSD objective by gradient descent.

\subsection{Farthest-point ensemble construction}
\label{SIsec:farthest-point_ensemble}

To construct ensembles of representative structures, we employed a farthest-point algorithm based on structural dissimilarity. The first structure was chosen as the frame with the largest RMSD to the mean conformation of the aligned trajectory (equivalently, the frame furthest from the centroid in the $3N_{\mathrm{beads}}$-dimensional coordinate space). We then iteratively added the structure with the largest minimum RMSD to all members of the current ensemble, where each RMSD was minimized over rigid-body rotations (see Algorithm \ref{SIalg:farthest-point_ensemble}):
\begin{equation}
	x_{\mathrm{new}}
	=
	\arg\max_{x \in \mathrm{trajectory}}
	\min_{x_m \in \mathrm{ensemble}}
	\mathrm{RMSD}_{\mathrm{align}}(x, x_m).
\end{equation}
This procedure promotes coverage of conformational space while avoiding redundancy. Ensembles of size $M$ were obtained by truncating this sequence, up to the prescribed maximum of $M_\mathrm{max}=100$.

\begin{algorithm}
\caption{Greedy farthest-point ensemble selection}
\label{SIalg:farthest-point_ensemble}
\begin{algorithmic}[1]

\State \textbf{Input:} structures $\{x_j\}_{j=1}^{N_{\mathrm{frames}}}$, maximum ensemble size $M_{\max}$
\State Compute the mean conformation $\bar{x} = \frac{1}{N_{\mathrm{frames}}} \sum_{j=1}^{N_{\mathrm{frames}}} x_j$.
\State Choose the first structure as the frame furthest from $\bar{x}$:
\[
j_1 = \operatorname*{arg\,max}_{j}\; \|x_j - \bar{x}\|.
\]
\State Initialize ensemble $\mathcal{E}_1=\{j_1\}$.
\State Initialize stored aligned distances $\widetilde D_{ij}=+\infty$.

\For{$M=1,\dots,M_{\max}-1$}
    \State Let $j_M$ be the most recently added structure.
    \State Compute aligned distances from $j_M$ to all trajectory structures:
    \[
    \widetilde D_{j_M,k}
    =
    \min_{R\in SO(3)}
    \mathrm{RMSD}(R\,x_{j_M},x_k),
    \qquad
    k \in \{1,\dots,N_{\mathrm{frames}}\}\setminus\mathcal{E}_M.
    \]
    \State Store these distances for reuse.
    \State For each candidate structure $k$, compute its distance to the current ensemble:
    \[
    d_{\min}(k)
    =
    \min_{j\in \mathcal{E}_M}
    \widetilde D_{j,k}.
    \]
    \State Select the structure farthest from the current ensemble:
    \[
    j_{M+1}
    =
    \operatorname*{arg\,max}_{k}
    d_{\min}(k).
    \]
    \State Update the ensemble:
    \[
    \mathcal{E}_{M+1}
    =
    \mathcal{E}_M \cup \{j_{M+1}\}.
    \]
\EndFor

\State \textbf{Output:} nested ensembles $\mathcal{E}_M$, $M=1,\dots,M_{\max}$

\end{algorithmic}
\end{algorithm}

\subsection{Hierarchical k-medoids ensemble construction}
\label{SIsec:Kmeds_ensemble}

As an alternative to the farthest-point method, we constructed
ensembles via a hierarchical k-medoids clustering procedure.
First, we ran K-medoids with $K = M_{\max}$ on the pre-aligned trajectory
(Euclidean distance in $3N_{\mathrm{beads}}$ dimensions) and retained the cluster medoids as candidates.

Then, to compare ensembles across all sizes $M \le M_{\max}$, we arranged the candidates to
form a nested sequence. In particular, we ordered them greedily by the facility-location (k-median) objective,
seeding with the overall candidate medoid and then at each step selecting
\begin{equation}
    c^* = \operatorname*{arg\,min}_{c \notin \mathcal{E}_k}
    \sum_{c'=1}^{M_{\max}}
    \min \bigl(d(c',\mathcal{E}_k),\;d_{cc'}\bigr),
    \label{SIeq:facility_location}
\end{equation}
where $d(c',\mathcal{E}_k)=\min_{j\in\mathcal{E}_k}d_{c'j}$.
Unlike the farthest-point rule (Section~\ref{SIsec:farthest-point_ensemble}),
which maximizes the maximum distance to the nearest representative,
this minimizes the sum of those distances. 
This resulted in a hierarchy of nested ensembles as in the farthest-point algorithm above, 
where ensembles of size $M' > M$ contain all members of the size-$M$ ensemble.

\subsection{Pre-noise imaging model}

Noiseless template images were generated using cryoJAX \cite{obrien_cryojax_2026}.
Because the molecular conformations come from a coarse-grained Martini simulation,
each effective ``atom'' is a CG bead.
We modeled each bead~$i$ as a 3D isotropic Gaussian density with standard deviation
$\sigma_{\mathrm{bead}} = 3.0$~\AA.
Setting the imaging axis to $z$ without loss of generality, integrating this density
over $z$ yields a noiseless projected template
\begin{equation}
    T(\bm{r})
    =
    \frac{1}{2\pi\sigma_{\mathrm{bead}}^2}
    \sum_{i=1}^{N_{\mathrm{beads}}}
    \exp\!\left(
        -\frac{|\bm{r}-\bm{r}_i|^2}{2\,\sigma_{\mathrm{bead}}^2}
    \right),
    \label{SIeq:template}
\end{equation}
where $\bm{r}=(r_x,r_y)$ is the 2D image coordinate and $\bm{r}_i$ are the
projected $xy$-coordinates of bead~$i$ after applying rotation~$R$.
After applying the CTF and translating by $\bm\tau$, the result is the full
noiseless template denoted $T_{x,R,\bm\tau}$ in the sections below.

The projected image is then multiplied in Fourier space by the contrast transfer
function (CTF) $H(\mathbf{k};\Delta f)$, which depends on the spatial
frequency~$\mathbf{k}$ and defocus~$\Delta f$.
The CTF is computed by cryoJAX using standard cryo-EM conventions at an accelerating
voltage of $300$~kV, for a specified defocus $\Delta f$.

Finally, the CTF-filtered template undergoes a uniform random in-plane translation
$\bm{\tau}=(\tau_x,\tau_y)$ drawn uniformly within a disk of radius
$r_{\mathrm{trans}}=20$~pixels, modeling the unknown position of the molecule within
the field of view.
Images are discretized onto a $128\times128$ pixel grid with spacing $3.0$~\AA/pixel.

Two approximations are implicit in the translation model.
First, we assume the molecule remains entirely within the field of view for all
translations in the prior support.
Under this condition, the signal is negligible near the image boundary, and each
translation can be implemented as a cyclic (wrap-around) shift, which admits an
efficient FFT-based cross-correlation.
If signal were to leak to the boundary, cyclic addressing would alias image content
across the edge and corrupt the likelihood.
The image size was chosen so that the distance from the projected molecular extent to
the image edge exceeds $r_{\mathrm{trans}}$ for all conformations in the ensemble.
Second, the continuous uniform prior over the translation disk is approximated by a
uniform sum over the discrete set of integer-pixel translations
$\{(\tau_x,\tau_y)\in\mathbb{Z}^2 : \tau_x^2+\tau_y^2\leq r_{\mathrm{trans}}^2\}$,
comprising $N_\tau$ grid points.

\subsection{Rotation sampling}

Rotations were parameterized by a viewing axis $\vec n$ and an in-plane rotation $\psi$. Imaging directions were generated by sampling points on the unit sphere using a Fibonacci lattice. Projection along the $z$-axis causes the transformation $(\vec n, \psi) \rightarrow (-\vec n, \psi+\pi)$ to produce identical images. Therefore, only viewing axes with positive $z$ component were retained.

Then, for each viewing axis, a set of uniformly spaced in-plane rotations $\psi$ was applied. The total number of sampled rotations was $N_R = N_{\vec{n}} \times N_{\psi}$, where $N_{\vec{n}}$ is the number of imaging directions and $N_{\psi}$ is the number of in-plane rotations per direction. The number of viewing angles was $N_{\vec{n}} = N_{\psi}^2$, reflecting the difference in dimensionality between the two contributions, so $N_R = N_{\psi}^3$. Using $N_{\psi} = 11$, $N_R = 1,331$.

\subsection{Noise level and marginalization of the likelihood function}
\label{SIsec:noise_level}

Independent Gaussian noise with variance $\signoise^2$ was added to each pixel to generate the observed image $y$. Noise levels in experiments are typically unknown and vary from image to image due to variations in ice thickness and other reasons \cite{baxter_determination_2009}. Furthermore, noise levels are typically  reported as a signal-to-noise ratio (SNR) \cite{seitz_simulation_2019}. Since SNR depends on the image support over which signal power is computed, its numerical value is convention-dependent. Here, signal power $\sigsignal^2$ was computed as the mean squared pixel value over pixels whose squared intensity exceeded 10\% of the maximum squared intensity, averaged over templates of $N_\mathrm{rot}$ rotations of the conformations in the farthest-point ensemble with $M = 100$. Then, $\mathrm{SNR}\equiv \sigsignal^2 / \signoise^2$, and we sampled $\signoise$ such that the SNR was log-uniform distributed in an experimentally-plausible range $\SNR\in [0.05,\,0.1]$ \cite{baxter_determination_2009, seitz_simulation_2019}. This corresponds to a prior distribution over noise levels of:
\begin{equation}
    P(\signoise^2) = \frac{1}{\signoise^2} \frac{1}{\log(\sigma_{\mathrm{max}}^2) - \log(\sigma_{\mathrm{min}}^2)}
\end{equation}
with $\sigma_{\mathrm{min}}$ and $\sigma_{\mathrm{max}}$ determined by the signal power and the SNR.

Thus, for given conformation, rotation, defocus, translation, and noise level,
\begin{equation}
	P(y \mid x, R, \bm\tau, \Delta f, \signoise^2) = \mathcal{N}(y; T_{x,R,\bm\tau}, \signoise^2 \, \I),
\end{equation}
where $T_{x,R,\bm\tau}$ is the noiseless template defined above. To marginalize over noise levels, we used the Laplace approximation, as in \cite{cossio_bayesian_2013}. Changing parameters to $v = \log(\signoise^2) \in [v_\mathrm{min}, v_\mathrm{max}]$, the marginal over noise levels can be written:
\begin{equation}
    \begin{aligned}
        P(y \mid x, R, \bm\tau, \Delta f) &= \int P(y \mid x, R, \bm\tau, \Delta f, \signoise^2) \, P(\signoise^2) \, \dd\signoise^2
        \\
        &= \int P(y \mid x, R, \bm\tau, \Delta f, v) \, P(v) \, \dd v
        \\
        &= \frac{1}{v_{\mathrm{max}} - v_{\mathrm{min}}}\, \frac{1}{(2\pi)^{(N_\mathrm{pix}/2)}} \int \exp\left(-\frac12\,\frac{\|y-T_{x,R,\bm\tau}\|^2}{\exp(v)} - \frac{N_\mathrm{pix}}{2} \, v \right) \, \dd v
    \end{aligned}    
\end{equation}
To maximize the argument of the exponential,
\begin{equation}
    \frac{\dd}{\dd v}\left(\frac12 \|y-T_{x,R,\bm\tau}\|^2 \, \exp(-v) + \frac12 N_\mathrm{pix} \, v\right) \Big|_{v=v^*}
    = -\frac12 \|y-T_{x,R,\bm\tau}\|^2\, \exp(-v^*) + \frac12 N_\mathrm{pix} = 0
\end{equation}
giving
\begin{equation}
    \exp(v^*) = (\signoise^2)^* = \frac{\|y-T_{x,R,\bm\tau}\|^2}{N_\mathrm{pix}}.
\end{equation}
In practice, if $v^*$ fell outside the range $[v_\mathrm{min}, v_\mathrm{max}]$, we set it to the nearest boundary value. 

The curvature at $v^*$ is
\begin{equation}
    \frac{\dd^2}{\dd v^2}\left(\frac12 \|y-T_{x,R,\bm\tau}\|^2 \, \exp(-v) + \frac12 N_\mathrm{pix} \, v\right) \Big|_{v=v^*}
    = \frac12 \|y-T_{x,R,\bm\tau}\|^2\, \exp(-v^*) = \frac{N_\mathrm{pix}}{2}
\end{equation}
is constant for all images $y$ and templates $T_{x,R,\bm\tau}$. Thus, the Laplace approximation to the marginalization over $\signoise^2$ is
\begin{equation}
    P(y \mid x, R, \bm\tau, \Delta f) \propto \mathcal{N}(y; T_{x,R,\bm\tau}, (\signoise^2)^* \, \I).
\end{equation}
Note that $(\signoise^2)^*$ depends on both $R$ and $\bm\tau$ through the residual
$\|y - T_{x,R,\bm\tau}\|^2$, so a separate Laplace approximation is applied for each
$(R, \bm\tau)$ pair.
The marginal likelihood $P(y \mid x)$ is then obtained by averaging over the joint
discrete rotation and translation grids:
\begin{equation}
    P(y \mid x, \Delta f) \approx \frac{1}{N_R\,N_\tau}
    \sum_{r=1}^{N_R} \sum_{t=1}^{N_\tau}
    P(y \mid x, R_r, \bm\tau_t, \Delta f).
    \label{SIeq:Pyx_joint}
\end{equation}

\subsection{Estimating conditional mutual information for cryo-EM data}
\label{SIsec:MI_estimation}

We estimate the conditional mutual information $\IayT$ using the Gaussian
approximation Eq.~\eqref{SIeq:IayT_eigs}.
This requires the $\theta$-averaged Fisher information
$\bar{\bm{F}}_{y\mid\alpha}(\alpha)$ for each $\alpha$, and its expectation over
$\alpha$ drawn from the Dirichlet prior.
Here $\theta_i \equiv \Delta f_i$ is the defocus of image $i$, which is measured
per image in a cryo-EM experiment and therefore treated as known.

\subsubsection{Image bank precomputation}

For each defocus value $\Delta f$, we precomputed a bank of noiseless template
images at zero translation: for each representative structure $x_m$ and each of
$N_R$ sampled rotations $R_r$, the template $T_{m,r}(\Delta f)$ was rendered using the forward model described in Section~\ref{SIsec:methods}.
Storing templates at $\bm\tau = 0$ is without loss of generality because
translations are applied during likelihood evaluation, as described below.
Templates are reshaped to long vectors of length $N_\mathrm{pix}=128^2$; their
squared norms $\|T_{m,r}(\Delta f)\|^2$ are precomputed since they appear in the
likelihood and are translation-invariant (translations preserve the $L^2$
norm).

Defocus values were drawn from a Gaussian distribution
$\Delta f \sim \mathcal{N}(\Delta f_0, \sigma_{\Delta f}^2)$ with
$\Delta f_0 = 17{,}500$~\AA\ and $\sigma_{\Delta f} = 2{,}500$~\AA.
To evaluate the expectation $\E_{\Delta f}[\bm{F}_{y|\alpha,\Delta f}(\alpha,\Delta f)]$, we used Gauss--Hermite (GH) quadrature
\cite{abramowitz_handbook_1964} with $N_{\Delta f} = 5$ nodes and weights.
The quadrature nodes are placed at
\begin{equation}
    \Delta f_i = \Delta f_0 + \sqrt{2}\,\sigma_{\Delta f}\,\xi_i,
    \label{SIeq:GH_nodes}
\end{equation}
where $\{\xi_i\}_{i=1}^{N_{\Delta f}}$ are the roots of the $N_{\Delta f}$-th
physicist's Hermite polynomial $H_{N_{\Delta f}}(\xi)$, and $\{w_i\}$ are the
corresponding GH weights normalized to sum to one.
The weights $w_i$ are the Gaussian probability mass at each node.

The key advantage of GH quadrature over Monte Carlo averaging over defocus is
that it is deterministic and exact for all polynomial integrands up to degree
$2N_{\Delta f}-1 = 9$: five nodes capture the smooth dependence of
$\bm{F}_{y|\alpha,\Delta f}(\alpha,\Delta f)$ on defocus without any additional Monte Carlo
variance.

\subsubsection{Likelihood evaluation}

For a fixed defocus $\Delta f_i$, the likelihood $P(y \mid x_{m'}, \Delta f_i)$
marginalizes over rotations and translations jointly following
Eq.~\eqref{SIeq:Pyx_joint}:
\begin{equation}
    P(y \mid x_{m'}, \Delta f_i)
    \approx
    \frac{1}{N_R\,N_\tau}
    \sum_{r=1}^{N_R}\sum_{t=1}^{N_\tau}
    P(y \mid x_{m'}, R_r, \bm\tau_t, \Delta f_i),
    \label{SIeq:Pyx_defocus}
\end{equation}
where each factor $P(y \mid x_{m'}, R_r, \bm\tau_t, \Delta f_i)$ uses the
Laplace approximation to marginalize over noise (Section~\ref{SIsec:methods}).
Translation shifts $\bm\tau_t$ are not applied to the stored templates; instead,
for each rotation~$r$, the cross-correlation $\langle y, T_{m',r,\bm\tau_t}\rangle$
for all $N_\tau$ translation grid points is evaluated simultaneously via FFT:
the 2D real FFTs of $y$ and of $T_{m',r}(\Delta f_i)$ are computed once, their
pointwise product taken in Fourier space, and the inverse FFT yields
$\langle y, T_{m',r,\bm\tau_t}\rangle$ at all integer-pixel translations at once.
This reduces the cost from $O(N_\tau \cdot N_\mathrm{pix})$ per rotation to
$O(N_\mathrm{pix}\log N_\mathrm{pix})$.
The log-sum-exp over all $N_R\times N_\tau$ pairs is accumulated in a single pass
to avoid storing the full $N_Y \times N_R \times N_\tau$ likelihood array.

\subsubsection{Fisher information estimation at fixed defocus}

For each defocus node $\Delta f_i$ and each $\alpha$ sample, we estimated the
Fisher information $\bm{F}_{y|\alpha,\Delta f_i}(\alpha,\Delta f_i)$ as follows.
We sampled a latent component index $m\sim\alpha$ and a rotation index $r$
uniformly from $\{1,\dots,N_R\}$, selecting a noiseless template
$T_{m,r}(\Delta f_i)$ from the bank.
A noise level $\signoise$ was drawn log-uniformly, and a noise vector
$z\sim\mathcal{N}(0,\I)$ was sampled; the observed image was set to
$y = T_{m,r}(\Delta f_i) + \signoise\,z$.
The per-image likelihood $P(y\mid x_{m'}, \Delta f_i)$ was computed for each
conformation $m'$ via Eq.~\eqref{SIeq:Pyx_defocus}, yielding the responsibility
vector
\begin{equation}
    r_m(y;\alpha,\Delta f_i)=\frac{P(y\mid x_m, \Delta f_i)\,\alpha_m}{P(y\mid\alpha, \Delta f_i)},
    \qquad
    P(y\mid\alpha, \Delta f_i) = \sum_{m=1}^M P(y\mid x_m, \Delta f_i)\,\alpha_m.
\end{equation}

The score in ambient $\alpha$-coordinates is $s_m(y;\alpha,\Delta f_i) = r_m(y;\alpha,\Delta f_i)/\alpha_m - 1$.
Averaging $N_Y = 100$ independent images yields the Monte Carlo estimate of the Fisher information for each defocus value:
\begin{equation}
    \bm{F}_{y|\alpha,\Delta f_i}(\alpha,\Delta f_i)
    \approx
    \frac{1}{N_Y}\sum_{j=1}^{N_Y}
    s(y_j;\alpha,\Delta f_i)\,s(y_j;\alpha,\Delta f_i)^\top.
\end{equation}

\subsubsection{Defocus averaging}

The Fisher information averaged over known, Gaussian-distributed defocus values (Eq.~\eqref{SIeq:Fisher_avg}) was approximated by
the Gauss--Hermite weighted sum,
\begin{equation}
    \bar{\bm{F}}_{y\mid\alpha}(\alpha)
    \approx
    \sum_{i=1}^{N_{\Delta f}} w_i\,\bm{F}_{y|\alpha,\Delta f_i}(\alpha,\Delta f_i).
    \label{SIeq:Fbar_quadrature}
\end{equation}
To reduce Monte Carlo variance, the same $\alpha$ sample and the same
random seeds for $(m, r, \signoise, z)$ were reused across all defocus values;
only the template bank changed with defocus.
This common-random-number coupling ensures that the weighted sum
$\bar{\bm{F}}_{y\mid\alpha}(\alpha)$ is estimated with much lower noise than
if independent samples were used for each defocus value.

\subsubsection{MI computation}

The Fisher information matrices above are in ambient $\alpha$-coordinates, but the mutual information depends on
its projection onto the tangent space of the simplex.
Therefore, for each $\alpha^{(a)} \sim \mathrm{Dir}(\beta\,\mathbf{1})$, 
we computed the projected Fisher information matrix,
\begin{equation}
	\tilde{\bar{\bm{F}}}_{y|\alpha}(\alpha) = \B^\top \bar{\bm{F}}_{y|\alpha}(\alpha) \, \B.
\end{equation}
Then we computed the $M-1$ eigenvalues of this matrix, $\bar\lambda_k(\alpha^{(a)})$.

Finally, the conditional mutual information was estimated by the
Monte Carlo average
\begin{equation}
    \IayT
    \approx
    \frac{1}{N_\alpha} \sum_{a=1}^{N_\alpha}
    \frac{1}{2} \sum_{k=1}^{M-1}
    \log\!\left(1 + \frac{N\,\bar\lambda_k(\alpha^{(a)})}{M(M\beta+1)}\right),
    \label{SIeq:IayT_MC}
\end{equation}
which is the sample-mean approximation to the expectation in
Eq.~\eqref{SIeq:IayT_eigs}.
Monte Carlo calculations were organized into $N_\mathrm{blocks} = 5$ independent
blocks of $N_\alpha = 20$ Dirichlet samples each; reported uncertainties are
standard errors across block estimates.

\subsubsection{Variance reduction across ensemble sizes}

To reduce variance when comparing different ensemble sizes $M$, we used common
Gamma random variables for Dirichlet sampling: within each block, all values of
$M$ used the same bank of pre-drawn Gamma variates.
Specifically, in each block we drew
\begin{equation}
    g_m^{(a)} \sim \mathrm{Gamma}(\beta,1),
    \qquad m=1,\dots,M_{\max},
    \qquad a=1,\dots,N_\alpha,
\end{equation}
and for an ensemble of size $M$, the Dirichlet sample was
\begin{equation}
    \alpha_m^{(a,M)}
    =
    \frac{g_m^{(a)}}{\sum_{j=1}^M g_j^{(a)}},
    \qquad m=1,\dots,M.
\end{equation}
Because ensembles with $M' < M$ are nested prefixes of the $M$-structure ensemble, the same image bank serves all $M$, and the shared Gamma variates reduce Monte Carlo noise in comparisons of $\IayT$ across $M$.

\section{Maximum-entropy reweighting and its Bayesian interpretation}
\label{sec:maxent_bayes}

Here we show that Maximum-entropy (MaxEnt) ensemble reweighting \cite{orioli_how_2020, borthakur_determining_2025} can be framed as Bayesian inference. Let $\alpha^0$ denote a reference set of weights, for example uniform weights assigned to all frames of a simulation. Suppose further that experimental data constrain a set of $K$ ensemble-averaged observables,
\begin{equation}
\bar O_j(\alpha)
=
\sum_{m=1}^M \alpha_m \, O_j(x_m),
\qquad j=1,\ldots,K ,
\end{equation}
where $O_j(x_m)$ is the value of observable $j$ in conformation $x_m$.

MaxEnt reweighting chooses the weights $\alpha$ that deviate as little as possible from the reference weights $\alpha^0$ while satisfying the experimental constraints \cite{bonomi_principles_2017, kofinger_efficient_2019, bottaro_integrating_2020, orioli_how_2020}. The entropy functional is
\begin{equation}
S(\alpha)
=
-\sum_{m=1}^M \alpha_m \, 
\log \frac{\alpha_m}{\alpha^0_m},
\end{equation}
so maximizing $S(\alpha)$ is equivalent to minimizing the Kullback--Leibler divergence
\begin{equation}
D_{\mathrm{KL}}(\alpha \| \alpha^0)
=
\sum_{m=1}^M \alpha_m
\log \frac{\alpha_m}{\alpha^0_m}.
\end{equation}
With hard experimental constraints,
\begin{equation}
\bar O_j(\alpha)
=
O_j^{\mathrm{exp}},
\qquad j=1,\ldots,K ,
\end{equation}
the constrained optimization problem is
\begin{equation}
\alpha^{\mathrm{ME}}
=
\arg\min_{\alpha}
D_{\mathrm{KL}}(\alpha \| \alpha^0),
\end{equation}
subject to the normalization and observable constraints. Introducing Lagrange multipliers $\lambda_j$ for the observable constraints and $\lambda_0$ for normalization gives
\begin{equation}
\mathcal L(\alpha,\lambda)
=
-\sum_{m=1}^M \alpha_m
\log \frac{\alpha_m}{\alpha^0_m}
- \lambda_0 \left(\sum_{m=1}^M \alpha_m - 1\right)
- \sum_{j=1}^K \lambda_j
\left(
\sum_{m=1}^M \alpha_m \, O_j(x_m)
-
O_j^{\mathrm{exp}}
\right).
\end{equation}
Optimizing over $\alpha_m$ gives the exponentially tilted weights
\begin{equation}
\alpha_m(\lambda)
=
\frac{
\alpha_m^0
\exp\!\left[-\sum_{j=1}^K \lambda_j \, O_j(x_m)\right]
}{
\sum_{n=1}^M
\alpha_n^0
\exp\!\left[-\sum_{j=1}^K \lambda_j \, O_j(x_n)\right]
} .
\label{eq:maxent_weights}
\end{equation}
The multipliers $\lambda_j$ are chosen so that the constraints $\bar O_j(\alpha)=O_j^{\mathrm{exp}}$ are satisfied.

The same construction can be relaxed to allow for noisy or uncertain experimental measurements. Suppose the experimental observables have uncertainty described by covariance matrix $\Sigma$, so that deviations between the measured observables and the ensemble predictions are penalized by
\begin{equation}
\chi^2(\alpha)
=
\left(O^{\mathrm{exp}}-\bar O(\alpha)\right)^T
\Sigma^{-1}
\left(O^{\mathrm{exp}}-\bar O(\alpha)\right).
\end{equation}
Then a soft-constraint maximum-entropy estimate can be written as
\begin{equation}
\alpha^{\mathrm{ME}}
=
\arg\min_{\alpha}
\left[
\frac{1}{2}\chi^2(\alpha)
+
\beta D_{\mathrm{KL}}(\alpha \| \alpha^0)
\right],
\label{eq:soft_maxent}
\end{equation}
again subject to $\alpha_m\geq 0$ and $\sum_m \alpha_m=1$. The parameter $\beta$ controls the strength of the regularization toward the reference ensemble. In the limit of very small experimental uncertainty, the soft constraint formulation approaches the hard constraint formulation.

This soft-constraint objective has a direct Bayesian interpretation. Let the likelihood for the experimental data be
\begin{equation}
P(O^{\mathrm{exp}} \mid \alpha)
\propto
\exp\!\left[
-\frac{1}{2}
\left(O^{\mathrm{exp}}-\bar O(\alpha)\right)^T
\Sigma^{-1}
\left(O^{\mathrm{exp}}-\bar O(\alpha)\right)
\right],
\end{equation}
and choose a prior over weights of the form
\begin{equation}
P(\alpha)
\propto
\exp\!\left[
-\beta D_{\mathrm{KL}}(\alpha \| \alpha^0)
\right],
\qquad
\alpha \in \Delta_{M-1},
\label{eq:entropic_prior}
\end{equation}
where $\Delta_{M-1}$ denotes the probability simplex. Bayes' rule gives
\begin{equation}
P(\alpha \mid O^{\mathrm{exp}})
\propto
P(O^{\mathrm{exp}} \mid \alpha) P(\alpha).
\end{equation}
Taking the negative logarithm of this posterior, up to constants independent of $\alpha$, gives exactly the objective in Eq.~\eqref{eq:soft_maxent}. Thus, soft maximum-entropy reweighting is the maximum a posteriori estimate of $\alpha$ under a Gaussian experimental noise model and an entropic prior centered on the reference weights $\alpha^0$.

Similarly, hard-constraint maximum-entropy reweighting can be viewed as the maximum a posteriori estimate obtained from a likelihood that enforces exact agreement with the experimental observables,
\begin{equation}
P(O^{\mathrm{exp}} \mid \alpha)
\propto
\prod_{j=1}^K
\delta\!\left(
O_j^{\mathrm{exp}}
-
\sum_{m=1}^M \alpha_m \, O_j(x_m)
\right),
\end{equation}
together with the same entropic prior in Eq.~\eqref{eq:entropic_prior}. In this case, the posterior is supported only on weights that exactly satisfy the experimental constraints, and the maximum a posteriori estimate is the constraint-satisfying $\alpha$ vector closest to $\alpha^0$ in relative entropy.

MaxEnt is one approach to solving the problem of inferring the weights $\alpha$, but it does not inform which ensemble $\{x_m\}$ to use for this inference problem. Furthermore, as a maximum a posteriori procedure, it does not by itself quantify posterior uncertainty or correlations among the weights. A fully Bayesian treatment would retain the full posterior distribution $P(\alpha \mid O^{\mathrm{exp}})$ rather than replacing it by a single point estimate.



\ifdefined\combinedwithmain\else
	\bibliographystyle{apsrev4-2}
	\bibliography{references}

\end{document}
\fi

	\bibliography{references}

\end{document}